\PassOptionsToPackage{prologue,usenames,dvipsnames,table}{xcolor}
\PassOptionsToPackage{bookmarks,unicode,colorlinks=true}{hyperref}
\newif\ifanonymous

\anonymousfalse

\ifanonymous
    \documentclass[sigconf,review,screen,anonymous]{acmart}
\else
    \documentclass[sigconf,screen]{acmart}
\fi

\usepackage{amssymb,amsmath,amsfonts}
\usepackage{thm-restate}
\usepackage[inline]{enumitem}
\usepackage{bm}
\usepackage{graphicx}
\usepackage{xspace}
\usepackage{adjustbox}
\usepackage{tikz}
\usepackage[normalem]{ulem}
\usepackage{booktabs}
\usepackage{algorithm}
\usepackage[noend]{algpseudocode}
\usepackage{subcaption}
\usepackage{natbib} 
\usepackage{tcolorbox}
\usepackage{booktabs}
\usepackage{multirow}
\usepackage{makecell}

\usepackage{thm-restate}

\usepackage[capitalize,nameinlink]{cleveref}
\crefname{lemma}{Lem.}{Lem.}
\crefname{example}{Exmp.}{Exmp.}
\crefname{section}{Sect.}{Sect.}
\crefname{appendix}{Appx.}{Appx.}
\crefname{problem}{Problem}{Problem}
\crefname{definition}{Def.}{Def.}
\crefname{theorem}{Thm.}{Thm.}
\crefname{corollary}{Cor.}{Cor.}
\crefname{algorithm}{Alg.}{Alg.}
\crefname{rule}{Rule}{Rule}

\newtheorem{theorem}{Theorem}
\newtheorem{definition}[theorem]{Definition}

\newtheorem{example}[theorem]{Example}
\newtheorem{remark}{Remark}
\newtheorem{proposition}[theorem]{Proposition}
\input{macros.tex}
\input{tikz.tex}
\usepackage{tikz}
\usetikzlibrary{automata,arrows,positioning}
\AtBeginDocument{%
  }

\setcopyright{acmlicensed}
\copyrightyear{2018}
\acmYear{2018}
\acmDOI{XXXXXXX.XXXXXXX}

\acmConference[Conference acronym 'XX]{Make sure to enter the correct
  conference title from your rights confirmation email}{June 03--05,
  2018}{Woodstock, NY}
\acmISBN{978-1-4503-XXXX-X/18/06}

\begin{document}
\title{Runtime Enforcement of CPS against Signal Temporal Logic}
\author{Han Su}
\email{suhan@ios.ac.cn}
\orcid{0000-0003-4260-8340}
\affiliation{%
  \institution{Institute of Software, CAS}
  \city{}
  \country{}
 }
 \affiliation{ 
  \institution{\& University of CAS}
  \city{Beijing}
  \country{China}
}

\author{Saumya Shankar}
\authornotemark[1]
\email{Saumya.shankar@auckland.ac.nz}
\orcid{0000-0002-1455-4106}
\affiliation{
  \department{Department of Electrical, Computer and Software Engineering}
  \institution{The University of Auckland}
  \city{Auckland}
  \country{New Zealand}
}

\author{Srinivas Pinisetty}
\email{spinisetty@iitbbs.ac.in}
\orcid{0000-0001-7779-8231}
\affiliation{
  \institution{Indian Institute of Technology}
  \department{School of Electrical Sciences}
  \city{Bhubaneswar}
  \country{India}
}

\author{Partha S. Roop}
\authornote{The corresponding authors}
\email{p.roop@auckland.ac.nz}
\orcid{0000-0001-9654-5678}
\affiliation{
  \institution{The University of Auckland}
  \department{Department of Electrical, Computer and Software Engineering}
  \city{Auckland}
  \country{New Zealand}
}

\author{Naijun Zhan}
\authornotemark[1]
\email{znj@ios.ac.cn}
\orcid{0000-0003-3298-3817}
\affiliation{
  \institution{Peking University}
  \department{School of Computer Science}
  \city{}
  \country{}
} 
\affiliation{
  \institution{\& Institute of Software, CAS}
  \city{Beijing}
  \country{China}
}
\renewcommand{\shortauthors}{H.~Su et al.}

\begin{abstract}



Cyber-Physical Systems (CPSs), especially those involving autonomy, need guarantees of their safety. Runtime Enforcement (RE) is a lightweight method to formally ensure that some specified properties are satisfied over the executions of the system.  Hence, there is recent interest in the RE of CPS. However, existing methods are not designed to tackle specifications suitable for the hybrid dynamics of CPS. With this in mind, we develop runtime enforcement of CPS using properties defined in Signal Temporal Logic (STL).
 
In this work, we aim to construct a runtime enforcer for a given STL formula to minimally modify a signal to satisfy the formula. To achieve this, the STL formula to be enforced is first translated into a timed transducer, while the signal to be corrected is encoded as timed words. We provide timed transducers for the temporal operators \emph{until} and \emph{release} noting that other temporal operators can be expressed using these two. Our approach enables effective enforcement of STL properties for CPS. A case study is provided to illustrate the approach and generate empirical evidence of its suitability for CPS.

\end{abstract}




\maketitle

\section{Introduction}






Modern Cyber-Physical Systems (CPSs), especially those with emerging AI-enabled modules, are becoming increasingly difficult to verify formally, and sometimes even impossible, due to the chaotic behavior of the AI modules. Runtime Enforcement (RE) \cite{schneider2000enforceable}, serving as a lightweight formal method, has attracted increasing research interest in recent years for the verification of CPSs.
In RE, an \emph{enforcer} is synthesized to monitor the executions of a black-box system at runtime, ensuring compliance with a set of desired properties. In the event of a violation, the enforcer employs evasive actions to output property complaint words.  
There are various evasive actions, such as:
\begin{enumerate*}[label=(\roman*)]
    \item blocking the execution \cite{10.1145/353323.353382},
    \item modifying the input sequence by suppressing and/or inserting actions \cite{10.1007/s10207-004-0046-8,10.1145/1455526.1455532}, and
    \item buffering input actions until they can be safely forwarded \cite{DBLP:journals/fmsd/FalconeMFR11,10.1007/978-3-030-32079-9_4,10.1007/978-3-642-35632-2_23,10.1007/s10703-014-0215-y,10.1016/j.scico.2016.02.008}.
\end{enumerate*}
However, these evasive actions may not be suitable for CPSs since delaying reactions or terminating the system may be impractical \cite{10.1145/3092282.3092291}. 

One key aspect of CPS is the need for active interaction between the controller (the Cyber part) and the plant (the Physical part) \cite{10.1145/3126500}. This interaction involves responding immediately to various events or signals emitted by the plant. Consequently, the enforcer must address erroneous executions in CPS without delay and ensure continuous operation. 

Different methods have been proposed to synthesize enforcers for CPSs. Authors in \cite{10.1007/978-3-662-46681-0_51} first introduced a framework to synthesize enforcers for reactive systems, focusing solely on safety properties and considering untimed properties expressed as automata. Subsequent studies, including \cite{10.1145/3126500,10.1145/3092282.3092291,10.1109/TII.2019.2945520}, extended this framework to include bi-directional runtime enforcement for CPSs. However, all these methods assume a system model in discrete time, meaning they presuppose that signals or events occur only at discrete `ticks'. Although this assumption simplifies the modeling, it is devoid of the expressive power of continuous time specifications. 


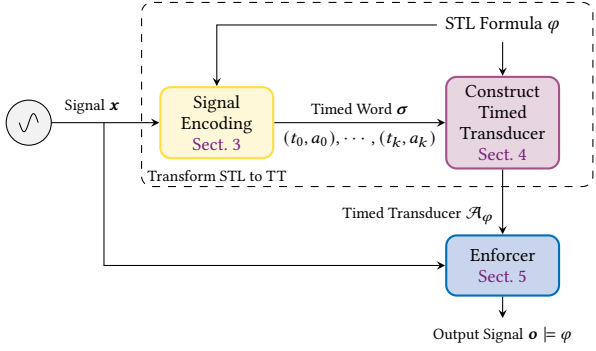
\begin{figure}[h]
    \centering
    \begin{adjustbox}{max width = 1\linewidth}
        \scalebox{1}{
            \begin{tikzpicture}[font=\footnotesize]
                \node (stl) at(3.8,1.3) {STL Formula $\varphi$};
                \node[sine wave icon, scale=0.4] (s) at (-2.5,0) {};
                \node[bubble=Goldenrod] (signal) at (0,0) {Signal\\Encoding\\\cref{sec: Signal Encoding}};
                \node[bubble=DarkOrchid] (trans) at (3.8,0) {Construct\\Timed\\Transducer\\\cref{sec: Transform STL into TA}};
                \node[bubble=NavyBlue] (modified) at (3.8,-1.9) {Enforcer\\\cref{sec:Runtime Enforcement against STL}};

                \draw[rounded corners, dashed] (-1, -0.85) -| (5,1.6) -- (-1,1.6) -- cycle;
                \node at(0,-0.7) {\scriptsize{Transform STL to TT}};
                
                \draw[-stealth] (s) -- node[above, xshift = -5pt]{\scriptsize{Signal $\signal$}} (signal);
                \draw[-stealth] (signal) -- node[below] {\scriptsize{$(t_0,a_0),\cdots,(t_k,a_k)$}} node[above]{\scriptsize{Timed Word $\tword$}} (trans);
                \draw[-stealth] (trans) -- node[left, yshift=-5pt]{\scriptsize{Timed Transducer $\automaton_\varphi$}} (modified);
                \draw[-stealth] (stl) -| (signal);
                \draw[-stealth] (stl) -- (trans);
                \draw[-stealth] (-1.5,0) |- (modified);
                \draw[-stealth] (modified) -- (3.8,-2.6) node[below] {\scriptsize{Output Signal $\resltsig\models\varphi$}};
            \end{tikzpicture}
        }
    \end{adjustbox}
    \vspace{-0.4cm}
    \caption{Overview}
    \label{fig:overview}
\end{figure}

In this paper, we take a step further by proposing a uniform hierarchy to synthesize enforcers for CPSs operating in dense time, with properties expressed using \emph{Signal Temporal Logic} (STL). The enforcer processes an input signal $\signal$ observed up to the current time $t$ and generates an output signal $\resltsig$ that satisfies the specified STL formula $\varphi$. As illustrated in \cref{fig:overview}, the enforcer encompasses three steps:
\begin{enumerate*}[label=(\roman*)]
\item Encoding a signal $\signal$ as a timed word in accordance with the STL formula. The timed word can be recognized by a timed automaton. This procedure is depicted as the yellow block in \cref{fig:overview}.
\item Constructing a variant of Timed Automaton (TA) from the given STL formula. The TA, with both input and output, is represented as \emph{Timed Transducer} (TT). The output of the TT indicates the enforcement strategy to be applied to the current input event (illustrated as the purple block in \cref{fig:overview}).
\item Enforcing a signal $\signal$ using the TT $\automaton_\varphi$ constructed from STL formula $\varphi$ (represented as the blue block in \cref{fig:overview}).
\end{enumerate*}
A significant advantage of our enforcer is that it \emph{does not} require reachability analysis because the enforcement strategies used are encoded explicitly within the timed transducer. Our experimental results demonstrate the effectiveness of our approach and provides empirical evidence of its suitability for CPS.

Our enforcer is characterized by \emph{soundness} (the output signal $\resltsig$ satisfy $\varphi$), \emph{transparency} (the input signal $\signal$ are only modified when necessary), and \emph{minimal modification} (the difference between the output and input signal values is minimal). Our work makes the following contributions:
\begin{enumerate}
    \item We propose a method to encode dense time signals into time words while preserving the information required to adjust the compliance of a given signal with a specified STL formula (\cref{sec: Signal Encoding}),
    \item we introduce a uniform approach to construct timed transducers against STL formulae, enabling these transducers to enforce the compliance of the STL formula on the input timed word (\cref{sec: Transform STL into TA}),
    \item we develop a method to minimally modify the signal to ensure its satisfaction with respect to the given STL formula (\cref{sec:Runtime Enforcement against STL}), and
    \item we provide experimental evidence to demonstrate the effectiveness of our approach (\cref{Case Study}).
\end{enumerate}

\paragraph{Organization} \cref{sec:Preliminaries and notations} provides a recap of important preliminaries and formally defines the problem. The process of encoding a signal into a timed word is detailed in \cref{sec: Signal Encoding}. The transformation of STL into TTs is discussed in \cref{sec: Transform STL into TA}. \cref{sec:Runtime Enforcement against STL} describes the method for signal modification and the comprehensive runtime enforcement algorithm for STL formulas. A relevant case study is presented in \cref{Case Study}. Related works are reviewed in \cref{sec:RL}. Finally, the conclusions and future work are outlined in \cref{sec:Conclusions and Future Works}. 

\section{Background and Problem Formulation}
\label{sec:Preliminaries and notations}
Let $\Nats$, $\Reals$, $\NonNegReals$, and $\NonNegRat$ denote the set of natural numbers, real numbers, non-negative real numbers and non-negative rational numbers, respectively. 
For a set $A \subseteq \mathbb{R}$ and a real number $a \in \mathbb{R}$, the expression $a \oplus A$ is used to denote the set obtained by adding $a$ to each element in $A$. 


    \subsection{Timed Transducer}\label{sec:Timed Transducer}
        A Timed Transducer (TT) is a specialized version of a timed automaton \cite{alur1994theory} that is capable of both taking input and producing output. We provide the essential preliminaries of timed transducers below.

        \paragraph{Timed Language}
        Let $\alphbt$ denote a finite alphabet. A pair $(t, a)\in\NonNegReals\times\alphbt$ is called an \emph{event}. A \emph{timed word} over $\alphbt$ is a finite sequence $\tword=(t_0,a_0)(t_1,a_1)\cdots(t_n,a_n)\allowbreak\in(\NonNegReals\times\alphbt)^*$, where $t_i$ is the \emph{time-stamp} indicating the global time at which the \emph{action} $a_i$ occurs, for all $0\le i \le n$. A \emph{timed language} $\mathcal{L}$ is a set of timed word, i.e., $\mathcal{L}\subseteq(\NonNegReals \times \alphbt)^*$. 

        
         \paragraph{Timed Transducer}
         Let $\clock$ be the set of clock variables. A \emph{clock constraint} $g$ is a Boolean combination of atomic constraints of the form $c\!\Join\! r$, with $c\in\clock$, $r\in \NonNegRat$, and $\Join\in\{\le,<,\ge,>,=\}$. We use $\mathcal{G}(\clock)$ to denote the set of clock constraints. A \emph{clock valuation} $v:\clock\mapsto\NonNegReals$ is a function assigning a non-negative real value to each clock $c\in\clock$. We write $v\models g$ if the clock valuation $v$ satisfies the clock constraints $g$. For $d\in\NonNegReals$, let $v+d$ denote the clock valuation which maps every clock $c\in\clock$ to the value $v(c)+d$, and for a set $\mathcal{C}'\subseteq\clock$, let $\clockreset{\mathcal{C}'}$ denote the clock valuation which resets all clock variables in $\mathcal{C}'$ to $0$ and agrees with $v$ for other clocks in $\clock\setminus\mathcal{C}'$. A timed transducer is defined as below:

        \begin{definition}[Timed transducer]
            A timed transducer is a tuple $\automaton=(\loca,\linit,\clock,\alphbt,\Lambda,\trans,\lambda,\lacpt)$, where
            \begin{itemize}
                \item $\loca$ is a finite set of locations;
                \item $\linit$ is the initial location;
                \item $\clock$ is the set of clocks;
                \item $\alphbt$ is the input alphabet;
                \item $\Lambda$ is the output alphabet;
                \item $\trans\subseteq \loca\times\alphbt\times\mathcal{G}(\clock)\times \power{\clock}\times\loca$ is a finite set of transitions;
                \item $\lambda: \trans \mapsto \Lambda$ is the output function that associates each transition with an output;
                \item $\lacpt\in\loca$ is a set of accepting locations;
            \end{itemize}  
        \end{definition}

        A transition \(\delta = (l, a, g, \clock', l')\) in \(\Delta\) represents a jump from location $l$ to $l'$ by performing an action \(a \in \alphbt\) when the constraint \(g \in \mathcal{G}(\clock)\) is satisfied by the current clock valuation. The set \(\clock'\) indicates which clocks should be reset upon reaching $l'$.

        A \emph{state} $q$ of \(\automaton\) is a pair $(l, v)$, where \(l \in \loca\) denotes the location, and $v$ is a clock valuation. A \emph{run} $\rho$ of \(\automaton\) over an input timed word \(\tword = (t_0, a_0)(t_1, a_1) \cdots (t_n, a_n)\) is a sequence \((l_0, v_0) \xrightarrow[b_0]{\tau_0, a_0} (l_1, v_1) \xrightarrow[b_1]{\tau_1, a_1} \cdots \xrightarrow[b_n]{\tau_n, a_n} (l_{n+1}, v_{n+1})\), where $\tau_i = t_i - t_{i-1}$ for $i = 1, 2, \ldots, n$ and $\tau_0 = t_0$, satisfying the following conditions:
        \begin{enumerate}
            \item $l_0$ is the initial location and $v_0(c) = 0$ for all $c\in\clock$,
            \item For each $i=0,1,\cdots,n$, there is a transition $\delta_i=(l_i,a_i,g_i,\clock_i,\allowbreak l_{i+1})\in\trans$ such that $v_i+\tau_i\models g_i$ and $v_{i+1} = \clockreset[(v_i+\tau_i)]{\clock_i}$,
            \item $\lambda(\delta_i) = b_i\in\Lambda$ for all $i=0,1,\cdots,n$.
        \end{enumerate}
        The run $\rho$ is \emph{accepted} by $\automaton$ if $l_{n+1}\in \lacpt$. The output timed word induced by $\automaton$ is $\bm{\omega}=(t_0,b_0)(t_1,b_1)\cdots(t_n,b_n)$, sharing the same timestamp $t_i$ ($i=0,1,\cdots,n$) as the input timed word. We use the notation \(\llangle\automaton\rrangle(\tword) = \bm{\omega}\) to denote that \(\automaton\) executes an accepted run over input timed word \(\tword\) that induces output timed word $\bm{\omega}$.

\begin{figure}[h]
		\centering
		\begin{tikzpicture}[->,shorten >=1pt,auto,node distance=2cm, el/.style = {inner sep=2pt, align=left, sloped},
		every label/.append style = {font=\small},
		semithick,initial where=below]
		
		\tikzstyle{every node}=[font=\small]
		\tikzstyle{good state}=[circle,thick,draw=NavyBlue!75,fill=NavyBlue!20,minimum size=5mm,accepting]
		\tikzstyle{bad state}=[circle,thick,draw=Maroon!75,fill=Maroon!20,minimum size=3mm]
		\node[initial,good state] (l0) {$l_0$};
		\node[good state]         (l1) [right of=l0, xshift = 15pt] {$l_1$};
		\node[bad state]       (l2) [right of=l1, xshift = 15pt] {$l_2$};
  
		\path (l0) edge [loop above] node {$w \mid \green{\top}$} (l0)
		edge node { $r,c:=0 \mid \green{\top}$ } (l1)             
		(l1) edge [loop above] node {$ w \mid \green{\top}$} (l1)
		edge [loop below] node {$r, c \geq 5, c:=0 \mid \green{\top}$} (l1)
        edge node {$r,c<5 \mid \green{\bot}$} (l2)
		(l2) edge [loop above] node {$ w \mid \green{\bot}$} (l2)
             edge [loop below] node {$r\mid \green{\bot}$} (l2);
		\end{tikzpicture}
		\caption{Timed Transducer $\mathcal{A}_P$}
		\label{fig:TA}
	\end{figure}
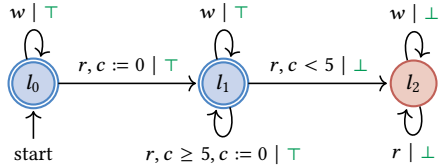

        \begin{example}\label{example:TA}
            The $\automaton_P$ illustrated in \cref{fig:TA} represents a TT for the property $P$: ``\textit{There should be a delay of at least $5$ time units between any two read file requests}''. This TT consists of locations $\loca=\{l_0,l_1,l_2\}$, with $l_0$ as the initial location and $\{l_0, l_1\}$ as the accepting locations, indicated by double circles. The input alphabet is $\alphbt=\{r, w\}$  with $r$ for read requests and $w$ for write requests. The output alphabet is $\{\top,\bot\}$ (denoted by green in \cref{fig:TA}), where $\top$ indicates a proper input that may lead to an accepted run, and $\bot$ indicates an improper input that leads to an unacceptable run. The transducer operates with one clock $c$.

            Given the input timed word $\tword=(1,r)(4,w)(6,r)$, the run $\rho$ of $\automaton$ progresses as follows:
            \[
                \rho = (l_0,0) \xrightarrow[\green{\top}]{1,r} (l_1,0) \xrightarrow[\green{\top}]{3,w} (l_1,3) \xrightarrow[\green{\top}]{2,r} (l_1,0).
            \]
            The output timed word of $\automaton_p$ over input timed word $\tword$ is \(\llangle\automaton_P\rrangle(\tword)=(1,\green{\top})(4,\green{\top})(6,\green{\top})\), which indicates whether the transition at current timestamp results in an acceptable run.
            \qedT
        \end{example}

    \subsection{Signal Temporal Logic}
    \label{sec:Signal Temporal Logic}
    Signal Temporal Logic (STL) \cite{maler2004monitoring} is a predicate logic used to describe and analyze continuous real-valued signals. Consider a signal $\signal:\NonNegReals\mapsto \Realn$. For each predicate $p(\signal)$, there is a corresponding function 
    $\mu_p:\Realn\mapsto \Reals$. The truth value of $p(\signal)$ at time $t$ is defined as follows:
    \begin{align*}
        p(\signal(t)) \Def\left\{
            \begin{aligned}
                \top,\quad \text{if}\quad \mu_p(\signal(t)) \ge 0,\\
                \bot,\quad \text{if}\quad \mu_p(\signal(t)) < 0.
            \end{aligned}\right.
    \end{align*}
    We will use the notion $p(\signal(t))\equiv\mu_p(\signal(t))\ge 0$ to define a predicate $p(x)$ in this paper. And we use \(|\signal|\) to denote the length of a signal.
    
    As demonstrated in \cite{fainekos2009robustness}, any STL formula can be equivalently converted into Negation Normal Form (NNF), in which negations appear only adjacent to predicates. In this paper, we considered \emph{non-nested} STL formula in NNF, which can be defined recursively as below:
        \begin{align*}
            &\phi \Def ~ \top ~\mid~ p(\signal) ~\mid~ \neg p(\signal) ~\mid~ \phi_1 \land \phi_2 ~\mid~ \phi_1 \lor \phi_2,\\ 
            &\varphi \Def ~ \phi_1 \until[I] \phi_2 ~\mid~ \phi_1 \release[I] \phi_2 ~\mid~ \varphi_1 \land \varphi_2 ~\mid~ \varphi_1 \lor \varphi_2~, 
        \end{align*}
        where $\until_I$ and $\release_I$ are the \emph{until} and \emph{release} operators, respectively. $I=[t_1,t_2]$ is a \emph{bounded} interval with \emph{rational} endpoints (i.e., $t_1,t_2\in\NonNegRat$)\footnote{The endpoints of $I$ are restricted to $\NonNegRat$ to facilitate encoding this into the clock constraints of the TT defined in \cref{sec:Timed Transducer}}. Note that $\release_I$ are the dual of $\until_I$, in a way that $\phi_1\release_I\phi_2\equiv\neg(\neg\phi_1\until_I\neg\phi_2)$. We use $pd(\varphi)$ to denote the set of \emph{predicates} in an STL formula $\varphi$. The NNF replaces the negation of a formula by including all operators and their duals in the grammar. Other temporal operators can be defined as syntactic sugars, e.g., 
            $\eventually_I\phi \equiv \top \until_I \phi,
            \always_I \phi \equiv \bot \release_I \phi$.
    \begin{remark}
        We employ NNF because it facilitates the subsequent transformation of an STL formula into a timed transducer. This choice is strategically important since timed transducer, acting as a special form of timed automata, are not closed under complementation. Thus, negating an STL formula would necessitate taking the complement of its corresponding timed automaton, which is problematic due to this lack of closure. 
    \end{remark}

    The semantics of STL is defined as the satisfaction of a formula $\varphi$ with respect to a signal $\signal$ and time $t \in \NonNegReals$. 
        
        \begin{definition}[STL Semantics]
        \label{def:STL Semantics}
        The satisfaction of an STL formula $\varphi$ at a given time $t$ over a signal $\signal$, denoted by $(\signal, t) \models \varphi$, is inductively defined as follows:
            \begin{align*}
                &(\signal, t) \models \top & & \\
                &(\signal, t) \models p(x) & &\tiff\quad \mu_p(\signal(t)) \ge 0\\
                &(\signal, t) \models \neg p(x) & &\tiff\quad \mu_p(\signal(t)) < 0\\
                &(\signal, t) \models \varphi_1 \land \varphi_2& & \tiff\quad  (\signal, t) \models \varphi_1 \aand (\signal, t) \models \varphi_2\\
                &(\signal, t) \models \varphi_1 \lor \varphi_2& & \tiff\quad  (\signal, t) \models \varphi_1 \text{ or } (\signal, t) \models \varphi_2\\
                &(\signal , t) \models \varphi_1 \until_I \varphi_2& & \tiff\quad \exists t' \in t \oplus I,~ (\signal, t') \models \varphi_2\\ 
                & && \qquad\quad\aand\forall t'' \in [t, t'], (\signal, t'') \models \varphi_1\\
                & (\signal, t) \models \varphi_1 \release_I \varphi_2& & \tiff\quad \forall t'\in t \oplus I,~ (\signal, t') \models \varphi_2\\
                & && \qquad\quad\oor \exists t''\in[t,t'], (\signal,t'')\models \varphi_1
            \end{align*} 
        \end{definition}
            Intuitively, the subscript $I$ in the until operator $\until_I$ defines the timing constraints under which a signal must \emph{eventually} satisfy $\varphi_2$, while ensuring that $\varphi_1$ is satisfied beforehand. Similarly, the subscript $I$ in the release operator $\release_I$ specifies the timing constraints in which a signal must \emph{always} satisfy $\varphi_2$, unless $\varphi_1$ has been satisfied earlier. We say  $\signal\models\varphi$ if $(\signal,0)\models \varphi$.
        
                \begin{example}
    		      \label{example:Properties in STL}
                The following examples illustrate some properties defined by STL.
    		    \begin{enumerate}
                    \item $(\signal  \leq 30) \until_{[5,10]} (\signal = 0)$: The value of the signal will be $0$ at a time instant between $5$ to $10$ seconds; until then the value of the signal is less than $30$.
    		        \item $\always_{[0,\infty)} (\signal < 3.5)$: The signal is always below $3.5$.
                    \item $\eventually_{[0,30]} (\signal > 100)$: At some time in the first $30$ seconds, the value of the signal will exceed $100$. \qedT
    		    \end{enumerate}
    	    \end{example}

    \subsection{Runtime Enforcement}
    \label{Preliminaries to Runtime Enforcement}
    The purpose of RE is to monitor input sequences produced by a running system and transform them into output sequences that adhere to a specified property $\varphi$. This is achieved using an enforcer.
    
    \paragraph{Constraints on an Enforcer.} Let $X$ denote the set of signals \(\signal : \NonNegReals \mapsto \Realn\). Some constraints are required on how enforcer $E_{\varphi}$ for $\varphi$ transforms a signal $\signal$ at time $t$, to ensure that it performs correctly and minimally disruptively. 

    \begin{definition}[Constraints on an Enforcer]\label{def:enforcer}
        Given an STL formula $\varphi$, an enforcer is a function $E_{\varphi}:X \mapsto X$ that satisfies the following conditions:
        \begin{itemize}
            \item \emph{Soundness}:
            \[
            \forall \signal \in X,~ E_\varphi(\signal)\models \varphi,
            \]
            \item \emph{Transparency}:
            \[
            \forall \signal \in X, ~ \signal\models\varphi \implies  E_\varphi(\signal) = \signal,
            \]
            \item \emph{Minimal Modification}:
            \[
            \forall \signal \in X,~ \signal\not\models\varphi \implies E_\varphi(\signal) = \argmin_{\resltsig\in O}||\signal-\resltsig||_s,
            \]
            where \(O = \{\resltsig \mid \resltsig \models \varphi \land |\signal| = |\resltsig|\}\), and $||\cdot||_s$ is the norm for signals defined as \(||\signal - \resltsig||_s \Def \max_t ||\signal(t) - \resltsig(t)||\), with $||\cdot||$ being the Euclidean norm in \(\Realn\).
        \end{itemize}
    \end{definition}
    

    

    \begin{figure*}[htbp]
            \centering
            \begin{adjustbox}{max width=1\linewidth}
                \begin{tikzpicture}[font=\small]
                    \centering
                    \begin{scope}
                        \draw[-Stealth, thick] (0,0) -- (6.3,0) node[right] {$t$};     
                        \draw[-Stealth, thick] (0,0) -- (0,1.7) node[above,left] {$x_1$};
                        \draw[dashed, color = gray] (0,1.05) node[left] {\black{$0.7$}} -- (6,1.05);
                        
                        \draw[color=NavyBlue, line width = 1.4pt, smooth, tension = 0.5] plot coordinates {
                            (0, 0.6)
                            (0.6, 1.05)
                            (1.2, 1.3)
                            (1.9, 1.2)
                            (2.7, 1.05)
                            (3.3, 0.94)
                            (3.8, 1.05)
                            (4.5, 1.3)
                            (5.4, 1.05)
                            (6.0, 0.8)
                        };
                        
                        \node at (0,0) [below] {\green{$0$}};
                        \draw[dashed, color = gray] (0.6, 0) node[below] {\orange{$0.5$}} -- (0.6,1.5);
                        \draw[dashed, color = gray] (2.7, 0) node[below] {\orange{$2.2$}} -- (2.7,1.5);
                        \draw[dashed, color = gray] (3.8, 0) node[below] {\orange{$3.2$}} -- (3.8,1.5);
                        \draw[dashed, color = gray] (4.8, 0) node[below] {\green{$4$}} -- (4.8,1.5);
                        \draw[dashed, color = gray] (5.4, 0) node[below] {\orange{$4.5$}} -- (5.4,1.5);
                        \draw[dashed, color = gray] (6,0) node[below] {\green{$5$}} -- (6,1.5);
                    \end{scope}

                    \begin{scope}[shift = {(8,0)}]
                        \draw[-Stealth, thick] (0,0) -- (6.3,0) node[right] {$t$};     
                        \draw[-Stealth, thick] (0,0) -- (0,1.7) node[above,left] {$x_2$};
                        \draw[dashed, color = gray] (0,0.75) node[left] {\black{$0.5$}} -- (6,0.75);

                        \draw[color=Maroon, line width = 1.4pt, smooth, tension = 0.5] plot coordinates {
                            (0, 1.2)
                            (1.4, 0.75)
                            (2.3, 0.15)
                            (3.5, 0.3)
                            (4.5, 0.6)
                            (5.6, 0.75)
                            (6.0, 1)
                        };

                        \node at (0,0) [below] {\green{$0$}};
                        \draw[dashed, color = gray] (1.4, 0) node[below] {\orange{$1.2$}} -- (1.4,1.5);
                        \draw[dashed, color = gray] (4.8, 0) node[below] {\green{$4$}} -- (4.8,1.5);
                        \draw[dashed, color = gray] (5.6, 0) node[below] {\orange{$4.7$}} -- (5.6,1.5);
                        \draw[dashed, color = gray] (6,0) node[below] {\green{$5$}} -- (6,1.5);
                    \end{scope}
                    
                    \node at(6,-1.3) {
                           $\begin{array}{c}
                                \textsf{timestamps}  \\
                                \textsf{actions}
                           \end{array}
                           \left(\!\!\!\begin{array}{c}
                               \green{0}  \\
                               \nblue{\neg p_1}\land \maroon{p_2} 
                          \end{array}\!\!\!\right) 
                          \left(\!\!\!\begin{array}{c}
                               \orange{0.5}  \\
                               \nblue{p_1}\land \maroon{p_2} 
                          \end{array}\!\!\!\right)
                          \left(\!\!\!\begin{array}{c}
                               \orange{1.2}  \\
                               \nblue{p_1}\land \maroon{\neg p_2} 
                          \end{array}\!\!\!\right)
                          \left(\!\!\!\begin{array}{c}
                               \orange{2.2}  \\
                               \nblue{\neg p_1}\land \maroon{\neg p_2} 
                          \end{array}\!\!\!\right)
                          \left(\!\!\!\begin{array}{c}
                               \orange{3.2}  \\
                               \nblue{p_1}\land \maroon{\neg p_2} 
                          \end{array}\!\!\!\right)
                          \left(\!\!\!\begin{array}{c}
                               \green{4}  \\
                               \nblue{p_1}\land \maroon{\neg p_2} 
                          \end{array}\!\!\!\right)
                          \left(\!\!\!\begin{array}{c}
                               \orange{4.5}  \\
                               \nblue{\neg p_1}\land \maroon{\neg p_2} 
                          \end{array}\!\!\!\right)
                          \left(\!\!\!\begin{array}{c}
                               \orange{4.7}  \\
                               \nblue{\neg p_1}\land \maroon{p_2} 
                          \end{array}\!\!\!\right)
                          \left(\!\!\!\begin{array}{c}
                               \green{5}  \\
                               \nblue{\neg p_1}\land \maroon{p_2} 
                          \end{array}\!\!\!\right)$
                    };
                \end{tikzpicture}
            \end{adjustbox}
            \caption{Signal Encoding against Formula $p_1\until_{[4,5]}p_2$}
            \label{fig:signal-encoding}
            \end{figure*}
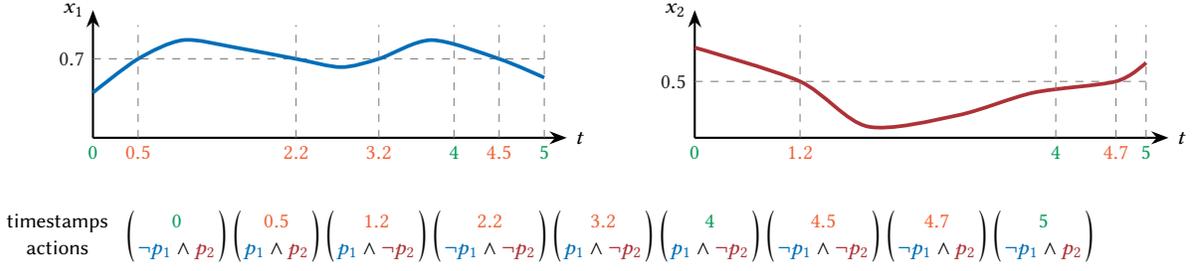
            
    Intuitively, \emph{soundness} ensures that the output signal complies with the specified STL formula $\varphi$. \emph{Transparency} stipulates that if the input signal $\signal$ already meets $\varphi$, the enforcer should not alter it, but rather transmit the original signal $\signal$ as the output. \emph{Minimal modification} requires that if the input signal $\signal$ does not satisfy $\varphi$, the enforcer should adjust it to ensure compliance with $\varphi$, while keeping the modifications as minimal as possible relative to the original signal $\signal$.
    


\paragraph{Problem Formulation}
\label{sec:Problem Formulation}
    With all the preliminary details established, we now formally define the problem under consideration in this paper as follows:
    \begin{tcolorbox}[boxrule=.5pt,colback=white,colframe=black!75]
        \textbf{Synthesis of Enforcer.} Given an STL formula $\varphi$, construct an enforcer $E_{\varphi} : X \mapsto X$ for $\varphi$ that satisfies the \emph{soundness}, \emph{transparency}, and \emph{minimal modification} conditions as per \cref{def:enforcer}.
    \end{tcolorbox}

\section{Signal Encoding}
\label{sec: Signal Encoding}

    In this section, we will introduce the procedure for encoding a signal into a timed word. This step is essential because we aim to enforce a signal using a TT, but a signal, defined as a real-valued function over dense time, is not directly compatible with TT. 
    
    The encoding process, applied to a given signal $\signal$ with respect to an STL formula $\varphi$, involves recording the truth value of predicates of $\varphi$ at both \emph{variable points} and \emph{relevant points} within the signal. We will now provide a detailed explanation of this encoding procedure.

    \paragraph{Variable Points}
    Intuitively, a variable point is where the truth value of a predicate regarding the signal changes.
        The concept of variable points is as below.
        \begin{definition}[Variable Point \customcite{bae2019bounded}{Def.~2.8}]
            Given a signal $\signal:\NonNegReals \mapsto \Realn$, a time point $\tau\in \NonNegReals$ is a variable point of $\signal$ with respect to a predicate $p(\signal)$ if for some neighborhood $B$ containing $\tau$, there are different truth values $u$ and $v$ such that $p(\signal)=u$ for every $t\in B\cap [0,\tau)$ and $p(\signal)=v$ for every $t\in B\cap (\tau,+\infty)$.
        \end{definition}

        In this paper, we limit our focus to \emph{non-Zeno} signals constrained within a \emph{bounded} time frame. Consequently, such signals possess a finite number of variable points. There is a notable characteristic of variable points, formalized in the following proposition:
        \begin{proposition}
            Given a signal $\signal$ and a predicate $p(\signal)$, assume $t_0 < t_1 < \cdots < t_k$ denote all the variable points of $\signal$ with respect to $p(\signal)$. It is then established that the truth value of $p(\signal)$ remains constant within each open interval $(t_i, t_{i+1})$ for all indices $i = 0, 1, \ldots, k$\footnote{Here, $t_{k+1}$ is defined to be the endpoint of the signal $\signal$.}.
        \end{proposition}

        Consequently, the signal $\signal$ can be effectively encoded as a timed word:
        \[
            \tword = (t_0, a_0), (t_1, a_1), \cdots, (t_k, a_k),
        \]
        where each $t_i$ denotes a variable point with respect to $p(\signal)$, and each $a_i$ represents the truth value of $p(\signal)$ within the interval $(t_i, t_{i+1})$ for indices $i = 0, 1, \ldots, k$. This encoding method ensures that $\tword$ comprehensively captures all information about $\signal$ with respect to the predicate $p(\signal)$. 

        The set of variable points of a signal with respect to an STL formula $\varphi$ is defined as the union of the sets of variable points for all predicates $p\in pd(\varphi)$\footnote{Recall that $pd(\varphi)$ is defined as the set of predicates in an STL formula $\varphi$ in \cref{sec:Preliminaries and notations}}. Accordingly, the timed word can be constructed based on these combined variable points. 
        The following example provides a comprehensive illustration of this process:

        \begin{example}\label{exp:variable-point}
            Consider the signal $\signal=(x_1, x_2)$ illustrated in \cref{fig:signal-encoding}, and the STL formula $\varphi = p_1 \until_{[4,5]} p_2$, with $p_1 \equiv x_1 - 0.7 \ge 0$, and $p_2 \equiv x_2 - 0.5\ge 0$. The time word for $\signal$ with respect to $\varphi$ is given by:
            \begin{align*}
            &(0.5,p_1\land p_2),(1.2, p_1 \land \neg p_2),(2.2, \neg p_1\land \neg p_2),(3.2, p_1\land \neg p_2),&&\\
            &(4.5,\neg p_1\land \neg p_2), (4.7, \neg p_1 \land p_2)
            &&\lhd
            \end{align*}
        \end{example}
        
        The values of a signal \(\signal\) that may lead to variable points against the predicate \(p(\signal)\) can be determined by solving the equation \(\mu_p(x)=0\). Depending on the structure of $\mu_p(x)$, different methods can be used to find or approximate the root of $\mu_p(x)$:
        \begin{enumerate*}[label=(\roman*)]
            \item Gaussian elimination is suitable for linear forms of $\mu_p(x)$,
            \item Newton-Raphsom method can be used for polynomial forms of $\mu_p(x)$,
            \item For more complex forms, such as transcendental equations, the Real roots isolation method \cite{gan2017reachability} can be used to approximate the intervals of the roots.
        \end{enumerate*}
        Using $vv(p)$ to denote the set of such valuations against predicate $p$, then $vv(\varphi) = \cup_{p\in pd(\varphi)}vv(p)$ for a given STL formula $\varphi$.  
        
        Variable points are effective for documenting changes in a signal in accordance with a specific STL formula. However, they may not provide sufficient information for \emph{enforcing} compliance with an STL property.
        Consider, for example, the scenario depicted in \cref{exp:variable-point}, where the proposition $p_1$ is false within the time interval $[0,0.5)$. This condition leads to the STL formula $\varphi$ not being satisfied by the signal $\signal$. However, in the absence of an input event before $t=0.5$ — specifically, an input at $t=0$ — no transducer can confirm that $\varphi$ is unsatisfied by the signal, nor can it enforce the signal accordingly.
        To address this, it becomes necessary to incorporate \emph{relevant points} tailored to an STL formula, ensuring all relevant events are recorded. 

        \paragraph{Relevant Points} Intuitively, relevant points include the time points that correspond to the interval boundaries of the given formula and the initial instant of the given signal. These points mark where the satisfaction requirements for predicates may change. For example, in the formula \( p_1 \until_{[t_1,t_2]} p_2 \), before \(t_1\), ensuring that \(p_1\) is satisfied suffices. However, after \(t_1\), it is also necessary to verify the satisfaction of \(p_2\).
        
        We proceed to inductively define relevant points of an STL formula as follows:
            \begin{definition}[Relevant Point]
                Given an STL formula $\varphi$, the set of relevant points $rp(\varphi)$ is inductively defined by:
                \begin{gather*}
                rp(\top) = \emptyset,\quad rp(\varphi_1\land\varphi_2) = rp(\varphi_1)\cup rp(\varphi_2),\\ 
                rp(p(x)) = \{ 0 \},\quad rp(\varphi_1\lor\varphi_2) = rp(\varphi_1)\cup rp(\varphi_2),\\
                rp(\varphi_1\until_{[t_1,t_2]}\varphi_2) = \{t_1,t_2\}\cup rp(\varphi_1) \cup rp(\varphi_2),\\
                rp(\varphi_1\release_{[t_1,t_2]}\varphi_2) = \{t_1,t_2\}\cup rp(\varphi_1)\cup rp(\varphi_2).
                \end{gather*}
            \end{definition}

            
            
            
            The actions of events at time points $t \in rp(\varphi)$ reflect the truth values of all predicates in $\varphi$ at time $t$, which can be directly sampled in real-time from the signal. Consider the following example for further illustration:
            
            \begin{example}\label{exp:relavant-point}
                Continuing \cref{exp:variable-point}, recall that $\varphi=p_1\until_{[4,5]}p_2$. The set of relevant point $rp(\varphi)=\{0,4,5\}$, and the events at relevant points with respect to signal $\signal=(x_1,x_2)$ is:
                \begin{align*}
                    (0,\neg p_1\land p_2),(4, p_1\land \neg p_2),(5,\neg p_1\land p_2).
                \end{align*}
                Consequently, the complete time word encoded from $\signal$ with respect to the STL formula $\varphi$ is depicted in \cref{fig:signal-encoding}. The events at variable points are highlighted in orange, while the events at relevant points are marked in green.
                \qedT
            \end{example}

        \paragraph{Signal encoding} We now give the process of signal encoding. 
        The signal encoding process, as outlined in \cref{alg:signal-encoding}, operates in real-time by monitoring changes in the truth values of predicates within the given STL formula. In an infinite loop, the algorithm waits for the input signal (\cref{line:await}) at current time $t$ (the function \textsf{current\_time()} can be used to get the current time as shown in \cref{line:time}). It updates the truth values \texttt{CurrPred} of all predicates (\cref{line:update value}) with respect to the current signal values. An event is emitted (\cref{line: emit}) whenever a variable point or relevant point is met (\cref{line:if condition}). 

    \begin{algorithm}[H]
        \caption{$\textsf{SignEncode}(\varphi)$}
        \label{alg:signal-encoding} 
        \begin{algorithmic}[1]
            \Require $\varphi$: STL formula
            \Ensure $\sigma$: time word encoded from $\signal$
            \State $\texttt{Rele} \gets rp(\varphi)$, $\texttt{Vari} \gets vv(\varphi)$,  $\texttt{Pred} \gets pd(\varphi)$ 
            \While {true}
                \State $\signal$ $\gets$ \textsf{await\_signal}() \label{line:await}
                \State $t$ $\gets$ \textsf{current\_time}() \label{line:time} \Comment{Get the current time $t$}
                \State $\texttt{CurrPred}\gets$ Truth values of predicates $p\in \texttt{Pred}$ with respect to $\signal$ at $t$\label{line:update value}
                \If{$\signal(t)\in\texttt{Vari}$ or $t\in\texttt{Rele}$}\label{line:if condition}
                    \State Emit $(t,\texttt{CurrPred})$ \label{line: emit}
                \EndIf
            \EndWhile
        \end{algorithmic}
    \end{algorithm}

    \begin{remark}
        During the signal encoding process, we can proactively identify variable points and relevant points, allowing for the enforcement of the signal before any actual violations occur. This proactive enforcement can be achieved by:
        \begin{enumerate*}[label=(\roman*)]
            \item expanding the values at a variable point into its surrounding neighborhood, and
            \item slightly adjusting the timing of relevant points in $rp(\varphi)$ to check these points in advance.
        \end{enumerate*}
    \end{remark}

\section{Constructing Timed Transducer from STL}
\label{sec: Transform STL into TA}

    \begin{figure*}[t]
        \begin{minipage}{.48\textwidth}
            \centering
            \begin{tikzpicture}[->,shorten >=1pt,auto,node distance=3.5cm,el/.style = {inner sep=2pt, align=left, sloped},every label/.append style = {font=\scriptsize},semithick,initial where=above]
			
			\tikzstyle{every node}=[font=\small]
			\tikzstyle{good state}=[circle,thick,draw=NavyBlue!75,fill=NavyBlue!20,minimum size=5mm,accepting]
			\tikzstyle{bad state}=[circle,thick,draw=Maroon!75,fill=Maroon!20,minimum size=3mm]
			\tikzstyle{dead state}=[rectangle,thick,draw=Maroon!75,fill=Maroon!20,minimum size=5mm]

                \node[initial, bad state]  at (0,-0.3) (S0) {$l_0$};
                \node[bad state] at (0,-2) (S2) {$l_1$};
			\node[good state] at (-2.5,-5.5)  (S3) {$l_2$};
			\node[bad state] at (2.5,-5.5)  (S4)  {$l_3$};
			
			\path (S0) edge node {$\begin{array}{c}
                                    p_1, c:=0, \mid \green{\top}\\ 
                                    \neg p_1, c:=0, \mid \green{\bot_{p_1}}  
                                    \end{array}$}(S2)
            
                      (S2) edge [loop right] node  {$\begin{array}{c} 
                                    p_1, c<t_1 \mid \green{\top} \\
                                    \neg p_1, c<t_1 \mid \green{\bot_{p_1}} \end{array}$} 
                      (S2) edge node [el,above] {$\begin{array}{c}
                                    p_1 \land p_2, c = t_1 \mid \green{\top}\\
                                    \neg p_1 \land p_2, c = t_1 \mid \green{\bot_{p_1}}\end{array}$} 
                      (S3) edge node [el,above] {$\begin{array}{c}
                                    p_1 \land \neg p_2, c=t_1 \mid \green{\top}\\
                                    \neg p_1 \land \neg p_2, c=t_1 \mid \green{\bot_{p_1}} \end{array}$} 
                      (S4)
			         (S3) edge [loop below] node {$\Sigma \mid  \green{\top}$} (S3)
            
                      (S4) edge [loop right] node [el,above] {$\begin{array}{c}
                                    p_1 \land \neg p_2, t_1 \le c < t_2 \mid \green{\top}\\ 
                                    \neg p_1 \land \neg p_2, t_1 \le c < t_2 \mid \green{\bot_{p_1}}  \end{array}$} 
                      (S4) edge node [el,above]{$\begin{array}{c} 
                                    p_1 \land p_2, t_1 \le c \leq t_2 \mid \green{\top}\\
                                    \neg p_1 \land p_2, t_1 \leq c \leq t_2 \mid \green{\bot_{p_1}}\end{array}$} (S3)
                      (S4) edge node {$\begin{array}{c}
                                    p_1 \land \neg p_2, c = t_2 \mid \green{\bot_{p_2}}\\
                                    \neg p_1 \land \neg p_2, c=t_2 \mid \green{\bot_{p_1}\land\bot_{p_2}}\end{array} $} (S3);
		\end{tikzpicture}
            \vspace{-.6cm}
            \caption{Timed Transducer $\automaton_{\until}$}
            \label{fig: until}
        \end{minipage}
        \hfill
        \begin{minipage}{.48\textwidth}
            \centering
            \begin{tikzpicture}[->,shorten >=1pt,auto,node distance=3.5cm,el/.style = {inner sep=2pt, align=left, sloped},every label/.append style = {font=\scriptsize},semithick,initial where=above]
			
			\tikzstyle{every node}=[font=\small]
			\tikzstyle{good state}=[circle,thick,draw=NavyBlue!75,fill=NavyBlue!20,minimum size=5mm,accepting]
			\tikzstyle{bad state}=[circle,thick,draw=Maroon!75,fill=Maroon!20,minimum size=3mm]
			\tikzstyle{dead state}=[rectangle,thick,draw=Maroon!75,fill=Maroon!20,minimum size=5mm]

                \node[initial, bad state]  at (0,-0.3) (S0) {$l_0$};
                \node[bad state] at (0,-2) (S1) {$l_1$};
                \node[good state] at (-2.5,-5.5) (S2) {$l_2$};
			\node[bad state] at (2.5,-5.5)  (S3) {$l_3$};

			\path (S0) edge node {$\neg p_1, c:=0 \mid \green{\top}$}
                      (S1) edge [bend right=40] node [el,above] {$ p_1 \mid \green{\top}$} (S2)
            
                      (S1) edge [loop right] node {$\neg p_1, c<t_1 \mid \green{\top}$} 
                      (S1) edge node [el,above] {$\begin{array}{c}
                                    \neg p_1 \land \neg p_2, c = t_1 \mid \green{\bot_{p_2}}\\ 
                                    \neg p_1 \land p_2, c = t_1 \mid \green{\top} \end{array}$} 
                      (S3) edge node [el,above] {$\begin{array}{c} 
                                    p_1, c < t_1 \mid \green{\top}\\
                                    p_1 \land p_2, c = t_1 \mid \green{\top}\\
                                    p_1 \land \neg p_2, c = t_1 \mid \green{\bot_{p_2}}\\ \end{array}$} (S2)
            
                      (S2) edge [loop below] node {$\Sigma \mid \green{\top}$} (S2)
            
                      (S3) edge [loop right] node [el,above]  {$\begin{array}{c}
                                    \neg p_1 \land \neg p_2, t_1 \le c < t_2 \mid \green{\bot_{p_2}}\\
                                    \neg p_1 \land p_2, t_1 \le c < t_2 \mid \green{\top}\end{array}$} 
                      (S3) edge node [el, above] {$\begin{array}{c}
                                    p_1 \land  p_2, t_1 \le c \leq t_2 \mid \green{\top}\\
                                    p_1 \land \neg p_2, t_1 \le c \leq t_2 \mid \green{\bot_{p_2}}\end{array}$} (S2)
                      (S3) edge node {$\begin{array}{c}
                                    \neg p_1 \land \neg p_2, c = t_2 \mid \green{\bot_{p_2}}\\
                                    \neg p_1 \land p_2, c = t_2 \mid \green{\top} \end{array}$} (S2);
		\end{tikzpicture}
            \vspace{-.6cm}
            \caption{Timed Transducer $\automaton_{\release}$}
            \label{fig: release}
        \end{minipage}

        \end{figure*}

    In this section, we outline a methodology for constructing a TT based on an STL. This TT processes the encoded timed word discussed in \cref{sec: Signal Encoding} and generates output for the corresponding enforcement strategy. 
    
    
    The construction process is inspired by the compositional hierarchy utilized in \cite{ferrere2019real} for building a TT for metric interval temporal logic (MITL). We have adopted this methodology and enhanced its applicability:
    \begin{enumerate*}[label=(\roman*)]
        \item It is suitable for STL, accommodating temporal operators with punctual intervals (e.g., \(\until_{[t_1,t_1]}\)),
        \item It is appropriate for runtime enforcement. The output of the TT serves as an enforcement strategy, which pinpoints the specific predicate causing the STL formula violation. This identification allows us to precisely modify only the signals involved in the failure, rather than altering all signals indiscriminately.
    \end{enumerate*}
    
    Initially, we will describe the construction of the TT for the temporal operators $\until_I$ and $\release_I$ used in the normal form of \cref{sec:Signal Temporal Logic}. Subsequently, we will present the method for composing these operators according to the structure of the STL formula.
    
    
    \subsection{Timed Transducer for $\until_I$ and $\release_I$}\label{subsec:transducer}

        \paragraph{TT for $p_1\until_{[t_1,t_2]}p_2$} We firstly present the construction of the TT \( \automaton_{\until} \), which is designed to enforce a signal according to the STL formula \( p_1 \until_{[t_1, t_2]} p_2 \). The structure of the transducer \( \automaton_{\until} \) is defined as follows:
        \begin{itemize}
            \item $\loca=\{l_0, l_1, l_2, l_3\}$; 
            \item $\linit=l_0$;
            \item $\clock=\{c\}$;
            \item $\alphbt= \{p_1,p_2\}$;
            \item $\Lambda = \{\top, \bot_{p_1}, \bot_{p_2} \}$;
            \item $\lacpt=\{l_2\}$.
        \end{itemize}
        
        The set of transitions $\Delta$ and the corresponding outputs $\lambda$ of \(\automaton_{\until}\) is depicted in \cref{fig: until}. 
        
        In the output alphabet, \(\top\) represents the strategy of making no change to the signal value, while \(\bot_{p_i}\) indicates that the signal value should be modified to satisfy the predicate \(p_i\). Essentially, an output of \(\bot_{p_i}\) suggests how the input action should be modified to achieve a \(\top\) output. For instance, if the transition from \(l_1\) to \(l_2\) in \cref{fig: until} has an input action of \(\neg p_1 \land p_2\) and outputs \(\bot_{p_1}\), it implies that the input should be changed to \(p_1 \land p_2\) to ensure a \(\top\) output in this transition.
        

        We propose below the equivalence of \emph{Until} operator of STL and its transducer. 
        \begin{restatable}{proposition}{restateUntil}
            \label{propo1}
            Let $\signal$ be a signal and $\tword$ denote its encoded timed word against the STL formula $p_1\until_{[t_1,t_2]}p_2$. Define $\bm{\omega}_{\top}$ as the timed word where all event actions are $\top$. The following equivalence is then established:
            \[
            \llangle\automaton_{\until}\rrangle(\tword) = \bm{\omega}_{\top} \iff \signal \models p_1 \until_{[t_1, t_2]} p_2
        \]
        \end{restatable}

        The proof relies on the case analysis based on the events received in time intervals $[0, t_1]$ and $(t_1,t_2]$. The detailed proof is provided in \cref{sec:appendix}.

        \paragraph{TT for $p_1\release_{[t_1,t_2]}p_2$}  Here, we detail the construction of the TT \( \automaton_{\release} \). The structure of the TT \( \automaton_{\release} \) is defined as follows:

        \begin{itemize}
            \item $\loca=\{l_0, l_1, l_2, l_3\}$; 
            \item $\linit=l_0$;
            \item $\clock=\{c\}$;
            \item $\alphbt=\{p_1,p_2\}$;
            \item $\Lambda = \{\top, \bot_{p_1}, \bot_{p_2} \}$;
            \item $\lacpt=\{l_2\}$.
        \end{itemize}
        where $\trans$ and the corresponding $\lambda$ are given in \cref{fig: release}.

         We propose below the equivalence of \emph{Release} operator of STL and its transducer. 

         \begin{restatable}{proposition}{restateRelease}
            \label{propo2}
            Let $\signal$ be a signal and $\tword$ denote its encoded timed word against the given STL formula $p_1\release_{[t_1,t_2]}p_2$. 
            Let $\bm{\omega}_{\top}$ be defined as before.
            The following equivalence is then established:
            \begin{align*}
            \llangle\automaton_{\release}\rrangle(\tword) = \bm{\omega}_{\top} \iff \signal \models p_1 \release_{[t_1, t_2]} p_2
            \end{align*}
        \end{restatable}
        
        The proof shares a similar idea with \cref{propo1} and is therefore omitted here.

        \begin{remark}
            Essentially, the TT we constructed is \emph{self-correcting}; that is, any transition within the TT has the potential to result in an acceptable run. This allows us to use such a TT to enforce a signal without worrying about the TT entering a violation state where no acceptable run exists.
        \end{remark}

    \subsection{Compositionally Constructing the Entire Timed Transducer}
        In this paper, because we consider non-nested STL, the possible connections between two sub-formulas containing temporal operators (i.e., \(p_1 \until_I p_2\) or \(p_1 \release_I p_2\)) are limited to either conjunction or disjunction. Consequently, it is sufficient to define the product of TTs we constructed in \cref{subsec:transducer} according to $\land$ or $\lor$.
        
        \paragraph{$\land$-Product} We will first explain how to construct TT of property in the form of $\varphi_1\land\varphi_2$ by taking product between TTs, where the TTs for $\varphi_1$ and $\varphi_2$ have been constructed as $\automaton_{1}$ and $\automaton_{2}$, respectively.
        
        \begin{definition}[$\land$-Product]
            Given two TTs $\automaton_1=(\loca_1,\linit^1,\clock_1,\alphbt_1,\Lambda_1,\trans_1,\allowbreak\lambda_1,\lacpt_1)$ and $\automaton_2=(\loca_2,\linit^2,\clock_2,\alphbt_2,\Lambda_2,\trans_2,\lambda_2,\lacpt_2)$\footnote{To avoid multiple subscripts, the indices of automata for the initial condition $l_0$ have been moved from superscript to subscript for both $\automaton_1$ and $\automaton_2$}, the $\land$-product automaton $\automaton_1\times_{\land}\automaton_2\Def (\loca,\linit,\clock,\alphbt,\Lambda,\trans,\lambda,\lacpt)$, where
            \begin{itemize}
                \item $\loca = \loca_1 \times \loca_2$,
                \item $\linit = (\linit^1,\linit^2)$,
                \item $\clock = \clock_1 \cup \clock_2$,
                \item $\alphbt = \alphbt_1 \cup \alphbt_2$,
                \item $\Lambda = \Lambda_1 \cup \Lambda_2$,
                \item $\delta = \left((l_1,l_2),(a_1,a_2),g_1 \land g_2,\clock_1'\cup\clock_2',(l_1',l_2')\right) \in \trans$ iff \\ 
                $\delta_1 = (l_1,a_1,g_1,\clock_1',l_1')\in \trans_1$ and $\delta_2 =(l_2,a_2,g_2,\clock_2',l_2')\in \trans_2$,
                \item $\lambda(\delta) = \lambda_1(\delta_1)\land\lambda_2(\delta_2) $,
                \item $\lacpt = \lacpt_1\times\lacpt_2$.
            \end{itemize}
        \end{definition}

        \paragraph{$\lor$-Product} We now explain how to construct TT of property in the form of $\varphi_1\lor\varphi_2$ by taking product between TTs, where the TTs for $\varphi_1$ and $\varphi_2$ have been constructed as $\automaton_1$ and $\automaton_2$, respectively.

        \begin{definition}[$\lor$-product]
            Given two TTs $\automaton_1=(\loca_1,\linit^1,\clock_1,\alphbt_1,\Lambda_1,\trans_1,\allowbreak\lambda_1,\lacpt_1)$ and $\automaton_2=(\loca_2,\linit^2,\clock_2,\alphbt_2,\Lambda_2,\trans_2,\lambda_2,\lacpt_2)$, the $\lor$-product automaton $\automaton_1\times_{\lor}\automaton_2\Def (\loca,\linit,\clock,\alphbt,\Lambda,\trans,\lambda,\lacpt)$, where
            \begin{itemize}
                \item $\loca = \loca_1 \times \loca_2$,
                \item $\linit = (\linit^1,\linit^2)$,
                \item $\clock = \clock_1 \cup \clock_2$,
                \item $\alphbt = \alphbt_1 \cup \alphbt_2$,
                \item $\Lambda = \Lambda_1 \cup \Lambda_2$,
                \item $\delta = \left((l_1,l_2),(a_1,a_2),g_1 \land g_2,\clock_1'\cup\clock_2',(l_1',l_2')\right) \in \trans$ iff \\ 
                $\delta_1 = (l_1,a_1,g_1,\clock_1',l_1')\in \trans_1$ and $\delta_2 =(l_2,a_2,g_2,\clock_2',l_2')\in \trans_2$,
                \item $\lambda(\delta) = \lambda_1(\delta_1)\lor\lambda_2(\delta_2) $,
                \item $\lacpt = (\lacpt_1\times\loca_2)\cup(\loca_1\times\lacpt_2)$.
            \end{itemize}
        \end{definition}
        
        Essentially, the primary distinction between the $\land$-product and the $\lor$-product lies in the output function $\lambda$ and the acceptance condition $F$. For a formula of the form $\varphi_1 \land \varphi_2$, the signal must satisfy both $\varphi_1$ and $\varphi_2$. Therefore, a transition in the product TT is considered `good' (i.e., the output action is $\top$) iff the transitions in both $\automaton_{\varphi_1}$ and $\automaton_{\varphi_2}$ are `good'. Additionally, the acceptance condition must ensure that the acceptance locations of both TTs are reached. Conversely, for a formula of the form $\varphi_1 \lor \varphi_2$, it is sufficient for the signal to satisfy either $\varphi_1$ or $\varphi_2$.
        
        By induction on the structure of a given STL formula, the following proposition holds:
        \begin{restatable}{proposition}{restateComposition}\label{prop:composition}
            Let $\signal$ be a signal and $\tword$ denote its encoded timed word against the given STL formula $\varphi_1\,op\,\varphi$, where $op\in\{\land,\lor\}$. Let $\bm{\omega}_{\top}$ be defined as in \cref{propo1}. The following equivalence is then established: 
            \[
                \llangle\automaton_{\varphi_1}\times_{op}\automaton_{\varphi_2}\rrangle(\tword) = \bm{\omega}_{\top} \iff \signal \models \varphi_1\,op\,\varphi_2.
            \]            
        \end{restatable}

\section{Runtime Enforcement using transducer}
\label{sec:Runtime Enforcement against STL}


    With all the preparatory work, we will present our runtime enforcement mechanism in this section. We will first describe an optimization-based method to \emph{minimally modify} the signal according to the output of the TT as detailed in \cref{sec: Transform STL into TA}. Then, we will introduce the enforcer against an STL formula, which will be outlined through a designated algorithm.

    \paragraph{Minimally modifying the signal}
    For a given STL formula $\varphi$, let \(x = \signal(t)\)\footnote{Note that the bold \(\signal\) represents a signal, while the non-bold $x$ is an $n$-dimensional real vector. We introduce the non-bold $x$ here to ease the symbolic burden in \cref{eq:opt}} be the value of the signal at the timestamp $t$. Assume the output action of TT \(\automaton_{\varphi}\) induced by the input $(t, a)$ is $\bot_{p_k}$. The signal can then be modified by solving the following optimization problem:
    \begin{align}\label{eq:opt}
        \begin{aligned}
        \text{Minimize:}  \quad & || y - x^{p_k} ||\\
        \text{Subject to:}\quad & \mu_i(x[x^{p_k}/y]) \ge 0, \quad \forall p_i\in a,\\
                                & \mu_j(x[x^{p_k}/y]) < 0, \quad \forall \neg p_j \in a,\\
                                & \mu_k(y) \Join 0,
        \end{aligned}
    \end{align}
    where \(\Join\) is $\ge$ if $\neg p_k \in a$ and \(\Join\) is $<$ otherwise. Here, $x^{p_k}$ denotes the components of $x$ related to the predicate $p_k$, $y$ is the decision variable of the optimization problem, representing the modified value of the signal at $t$. Notation $x[x^{p_k}/y]$ denotes the vector obtained by replacing the occurrences of $x^{p_k}$ in $x$ with $y$. For an intuitive illustration, see \cref{fig:illus}, where $y=(y_1,y_2,y_3)$.

    \begin{figure}[h]
        \centering
            \begin{align*}
                  & \quad x^{p_1} \Def \{x_1,x_2,x_m\}\,,\quad\ \,\cdots~ ,\quad \maroon{x^{p_k} \Def \{x_m,x_{m+1},x_n\}}\\
                  & p_1\equiv \mu_1(\tikzmarknode{x1}{x_1},\tikzmarknode{x2}{x_2},\tikzmarknode{xm}{x_m})\ge 0\,, ~\cdots ~,\, \maroon{p_k\equiv\mu_k(\tikzmarknode{xm1}{x_m},\tikzmarknode{xm11}{x_{m+1}},\tikzmarknode{xn}{x_n})\ge 0}\\
                  & \\
                x\quad\ =~& ~\left(~\tikzmarknode{p11}{x_1}\,,~\tikzmarknode{pk1}{x_2}\,,~x_3\,,~\cdots\,,~\tikzmarknode{p12}{\highlight{Maroon}{$x_m$}}\,,~\tikzmarknode{pk2}{\highlight{Maroon}{$x_{m+1}$}}\,,~\cdots\,, ~x_{n-1}\,,~\tikzmarknode{pk3}{\highlight{Maroon}{$x_n$}}~\right)\\
                x[x^{p_k}/y] = &~\left(~ x_1\,,~x_2\,,~ x_3\,,~\cdots\,,~\highlight{NavyBlue}{$y_1$\ }\,,~\highlight{NavyBlue}{\ \ $y_2$\ \ \,}\,,~\cdots\,,~x_{n-1}\,,~\highlight{NavyBlue}{$y_3$}~\right)
            \end{align*}
            \begin{tikzpicture}[overlay,remember picture,>=stealth,nodes={align=left,inner ysep=1pt}]
                \draw (p11.north)++(0,0.05) -- ++(0,0.6) -| (x1);
                \draw (pk1.north)++(0,0.05) -- ++(0,0.4) -| (x2);
                \draw (p12.north)++(-0.1,0) -- ++(0,0.37) -| (xm);

                \draw[Maroon] (p12.north)++(0.1,0) -- ++(0,0.57) -| (xm1.south);
                \draw[Maroon] (pk2.north) -- ++(0,0.37) -| (xm11.south);
                \draw[Maroon] (pk3.north) -- ++(0,0.37) -| (xn.south);
            \end{tikzpicture}
        \vspace{-5mm}
        \caption{Illustration of $x^{p_k}$ and $x[x^{p_k}/y]$.} 
        \label{fig:illus}
    \end{figure}


    We refer to this procedure as $\textsf{Modify}\left(x,a,b,\varphi\right)$, where $x$ represents the value of the signal, $a$ is the input action of the TT, $b$ is the corresponding output action, and $\varphi$ is the STL formula. The following proposition confirms the robustness of our minimal modification method:
    \begin{proposition}\label{prop:modify}
        If the optimization problem in \cref{eq:opt} is solvable, then this procedure maintains the minimal modification requirement as per \cref{def:enforcer}.
    \end{proposition}
    
    \begin{remark}
        \cref{eq:opt} can be solved using different methods, depending on the constraints provided in the predicate functions $\mu_p$ in the STL formula. If all of the $\mu_p$ are linear, then \cref{eq:opt} can be solved using Quadratic Programming (QP). If the $\mu_p$ are polynomial, \cref{eq:opt} can be transformed into Semidefinite Programming (SDP) by using Putinar's Positivstellensatz \cite{putinar1993positive}.  
    \end{remark}
   \paragraph{Enforcer} We are now ready to present our runtime enforcement algorithm, as shown in \cref{algorithm}. Assume we have a signal $\signal$ to be enforced against an STL formula $\varphi$. The algorithm begins by computing its TT $\automaton_{\varphi}$ following the method described in \cref{sec: Transform STL into TA} (\cref{al2line:0} in \cref{algorithm}). Subsequently, as the signal $\signal$ is received, it is encoded into input events. The enforcer $E_{\varphi}$ in \cref{algorithm} then traverses the TT $\automaton_{\varphi}$ and generates the output events. Depending on this output, the signal is modified (if required) and released.
   
    The algorithm proceeds as follows:
    \texttt{currState} monitors the current state of the TT, which includes the current location and clock valuation in the timed transducer. \texttt{currState} is initially set to the starting state of $\automaton_{\varphi}$ (\cref{al2line:1}). It then enters an infinite loop (\cref{al2line:2}) until an event is detected from \cref{alg:signal-encoding} (\cref{al2line:3}).

    Upon receiving an event, the transducer $\automaton_{\varphi}$ transitions, updates the \texttt{currState}, and gives the output $b$ according to the transition (\cref{al2line:4}). If the output is anything other than $\top$, the transducer minimally modifies the signal before it is released (\cref{al2line:6}).
    The following example illustrates how our enforcer operates. 

    \begin{figure*}[t]
            \begin{minipage}{.48\textwidth}
                    \centering
                    \begin{tabular}{c c c c c}
                        \toprule
                        \thead{state before\\transition} &\thead{timestamp} &  \thead{input\\action} & \thead{state after\\ transition} & \thead{output\\action}  \\
                        \midrule
                         $(l_0,0)$ & 0 & $\neg p_1\land p_2$ & $(l_1,0)$ & $\bot_1$ \\
                         $(l_1,0)$ & 0.5 & $ p_1\land p_2 $  & $(l_1,0.5)$ & $\top$   \\
                         $(l_1,0.5)$ & 1.2 & $ p_1\land\neg p_2 $ & $(l_1,1.2)$ & $\top$   \\
                         $(l_1,1.2)$ & 2.2 & $\neg p_1\land \neg p_2 $ & $(l_1,2.2)$ & $\bot_1$ \\
                         $(l_1,2.2)$ & 3.2 & $p_1\land \neg p_2 $     & $(l_1,3.2)$ & $\top$   \\
                         $(l_1,3.2)$ & 4   & $p_1\land \neg p_2 $     & $(l_3,4)$ & $\top$   \\
                         $(l_3,4)$ & 4.5 & $\neg p_1\land \neg p_2 $  & $(l_3,4.5)$ & $\bot_1$   \\
                         $(l_3,4.5)$ & 4.7 & $\neg p_1\land p_2 $     & $(l_2,4.7)$ & $\bot_1$\\  
                         \bottomrule
                    \end{tabular}
                    \captionof{table}{Transitions in TT of $p_1\until_{[4,5]} p_2$}
                    \label{tab:table_enf}
            \end{minipage}
            \hfill
            \begin{minipage}{.48\textwidth}
                \centering
                \vspace{-3mm}
                \begin{adjustbox}{max width = .95\linewidth}
                \begin{tikzpicture}[font=\small]
                    \centering
                    \begin{scope}
                        \draw[-Stealth, thick] (0,0) -- (6.3,0) node[right] {$t$};     
                        \draw[-Stealth, thick] (0,0) -- (0,1.7) node[above,left] {$x_1$};
                        \draw[dashed, color = gray] (0,1.05) node[left] {\black{$0.7$}} -- (6,1.05);
                        
                        \draw[color=NavyBlue, line width = 1.4pt, smooth, tension = 0.5] plot coordinates {
                            (0.6, 1.05)
                            (1.2, 1.3)
                            (1.9, 1.2)
                            (2.7, 1.05)};
                        \draw[color=NavyBlue, line width = 1.4pt, smooth, tension = 0.5] plot coordinates {
                            (3.8, 1.05)
                            (4.5, 1.3)
                            (5.4, 1.05)
                        };
                        \draw[color=red, line width = 1.6pt] plot coordinates{(0,1.05) (0.6,1.05)};
                        \draw[color=red, line width = 1.6pt] plot coordinates{(2.7,1.05) (3.8,1.05)};
                        \draw[color=red, line width = 1.6pt] plot coordinates{(5.4,1.05) (5.6,1.05)};
                        
                        \node at (0,0) [below] {\green{$0$}};
                        \draw[dashed, color = gray] (0.6, 0) node[below] {\orange{$0.5$}} -- (0.6,1.5);
                        \draw[dashed, color = gray] (2.7, 0) node[below] {\orange{$2.2$}} -- (2.7,1.5);
                        \draw[dashed, color = gray] (3.8, 0) node[below] {\orange{$3.2$}} -- (3.8,1.5);
                        \draw[dashed, color = gray] (4.8, 0) node[below] {\green{$4$}} -- (4.8,1.5);
                        \draw[dashed, color = gray] (5.4, 0) node[below] {\orange{$4.5$}} -- (5.4,1.5);
                        \draw[dashed, color = gray] (6,0) node[below] {\green{$5$}} -- (6,1.5);
                    \end{scope}

                    \begin{scope}[shift = {(0,-2.3)}]
                        \draw[-Stealth, thick] (0,0) -- (6.3,0) node[right] {$t$};     
                        \draw[-Stealth, thick] (0,0) -- (0,1.7) node[above,left] {$x_2$};
                        \draw[dashed, color = gray] (0,0.75) node[left] {\black{$0.5$}} -- (6,0.75);

                        \draw[color=Maroon, line width = 1.4pt, smooth, tension = 0.5] plot coordinates {
                            (0, 1.2)
                            (1.4, 0.75)
                            (2.3, 0.15)
                            (3.5, 0.3)
                            (4.5, 0.6)
                            (5.6, 0.75)
                            (6.0, 1)
                        };

                        \node at (0,0) [below] {\green{$0$}};
                        \draw[dashed, color = gray] (1.4, 0) node[below] {\orange{$1.2$}} -- (1.4,1.5);
                        \draw[dashed, color = gray] (4.8, 0) node[below] {\green{$4$}} -- (4.8,1.5);
                        \draw[dashed, color = gray] (5.6, 0) node[below] {\orange{$4.7$}} -- (5.6,3.8);
                        \draw[dashed, color = gray] (6,0) node[below] {\green{$5$}} -- (6,1.5);
                    \end{scope}
                     
                \end{tikzpicture}
                \end{adjustbox}
                \vspace{-4mm}
                \caption{Enforced Signal in \cref{exp:alg2}}
                \label{fig:after-enforce}
            \end{minipage}
        \end{figure*}
        
    \begin{algorithm}[H]
        \caption{Algorithm Enforcer $E_\varphi(\signal$) }
        \label{algorithm} 
        \begin{algorithmic}[1]
            \State $\automaton_\varphi \gets $ TT constructed from $\varphi$ \label{al2line:0}
            \State $ \texttt{currState} \leftarrow  [l_0 , c:=0]$\label{al2line:1}
            \While {$true$}\label{al2line:2}
                \State $(t, a)\gets$ event emitted by \cref{alg:signal-encoding}\label{al2line:3}
                \State $\texttt{currState}, ~b =\textsf{make\_transition}_{\automaton_{\varphi}}(\texttt{currState}, t, a)$\label{al2line:4}
                \If{$b \neq\top$}
                    \State $\signal(t)=\textsf{Modify}(\signal(t),a,b, \varphi)$\label{al2line:6}
                \EndIf
                \State release $\signal$
            \EndWhile
        \end{algorithmic}
    \end{algorithm}


  \begin{example}[Enforcement of STL formula on a timed word]\label{exp:alg2}
      Continuing to \cref{exp:relavant-point}, recall that the STL property is defined as $p_1\until_{[4,5]} p_2$, where $p_1\equiv x_1\ge 0.7$, $p_2\equiv x_2 \ge 0.5$. 
      \cref{tab:table_enf} gives the steps of enforcement of the timed word using Until Transducer. The signal at time points \{0, 2.2, 4.5, 4.7\} are modified to satisfy the STL formula. The modified signal is shown in \cref{fig:after-enforce}.
    \qedT
    \end{example}

    The following theorem states the correctness of the enforcer described in \cref{algorithm}
    \begin{theorem}
        Given an STL formula $\varphi$ and a signal $\signal$, the enforcer $E_\varphi$ in \cref{algorithm} can enforce $\signal$ to satisfy $\varphi$, while ensuring that the \emph{soundness}, \emph{transparency}, and \emph{minimal modification} conditions in \cref{def:enforcer} are met. 
    \end{theorem}
    \begin{proof}
    The transparency of the enforcer is a direct result of \cref{propo1}, \cref{propo2}, and \cref{prop:composition}, as \cref{algorithm} will not modify the signal under the $\top$ output of the TT. The soundness of the enforcer is ensured by observing the truth that, the $\bot_p$ outputs of TT essentially indicate how to modify the input action to those inputs that can lead to a $\top$ output. The minimal modification condition is ensured by \cref{prop:modify}.
    \end{proof}


    \paragraph{Complexity Analysis} The time complexity of \cref{algorithm} is multifaceted. The time complexity of the function $\textsf{make\_transition}_{\automaton{\varphi}}$ is $\mathcal{O}(m \times n)$, where $m$ is the number of states in the TT and $n$ is the size of the input alphabet. The time complexity of the function $\textsf{Modify}$ depends on the structure of the predicate function in the STL formula; it will be polynomial in the number of decision variables when the predicate functions are linear \cite{nesterov1994interior}.

    Other procedures, such as constructing the TT from the STL formula in \cref{sec: Transform STL into TA} (polynomial in the size of the TT, primarily influenced by the composition operator), and computing the values of the signal leading to the variable points in \cref{sec: Signal Encoding} (achieving quadratic convergence with the Newton-Raphson method), may be time-consuming. However, both procedures can be performed \emph{offline}, thus they do not impact the efficiency of our runtime enforcement algorithm.
\section{Case Study}
\label{Case Study}
We developed a prototype of our runtime enforcement algorithm in Python and applied this prototype to case studies on Autonomous Vehicles (AVs) to demonstrate the efficiency and scalability of our method. Three cases were considered in the experiments. The first case addresses the property of `safe stopping of AVs', the second focuses on `safe charging of AVs', while the third focuses on `safe deceleration of AVs'. All the cases underscore the efficiency (\cref{sec:efficiency}) and scalability (\cref{sec:scala}) of our method.



\subsection{Efficiency Evaluation}\label{sec:efficiency}

    \paragraph{Safe stopping of AVs}
    Consider a scenario in which an AV is required to decelerate to a complete stop when approaching a red light or a designated stop point. This requirement is expressed by the property \((v \le 30)\until_{[5,10]} (v=0)\). This stipulates that \textit{the speed of the vehicle must ultimately reach $0$ within a time frame of $5$ to $10$ seconds, while maintaining a speed no larger than $30$ until then}.

    \begin{figure}[h]
        \centering
        \includegraphics[width=.76\linewidth,trim={18cm 0cm 0cm 0cm}, clip]{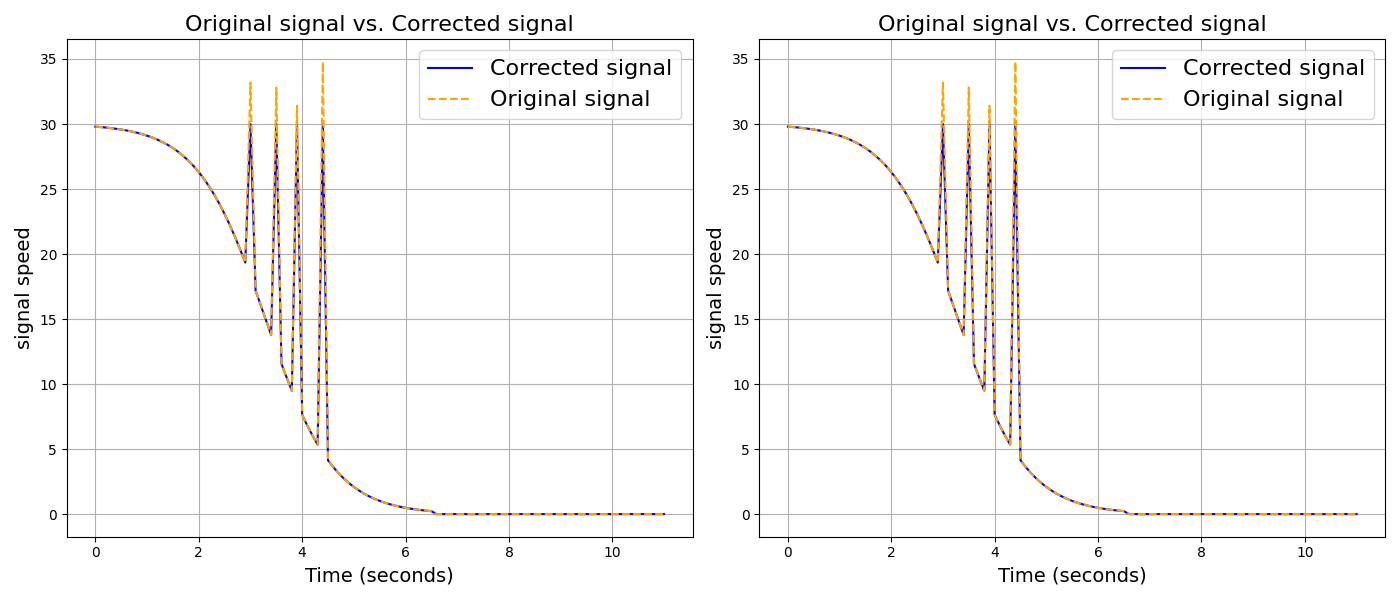}
        \caption{Enforcement of Speed Signal against Safe Stopping Property}
        \label{fig:speed}
    \end{figure}

    The results of our experiment are depicted in \cref{fig:speed}, where the blue signal represents the output after enforcement, while the orange one is the original signal. These results demonstrate that the enforcement monitor effectively adjusted the signal to ensure compliance with the STL property, while maintaining transparency and minimal modification. Specifically, the enforcer precisely addressed the four instances where the speed exceeded $30$ (sudden speed spikes), applying only the necessary changes without superfluous adjustments to the signal.
  
    \paragraph{Safe charging of AVs}
    Consider a scenario in the battery charging systems of AVs. Normally, the current stays within a safe range throughout a specified interval. If, however, the voltage reaches a specific volts, the system switches to a charging mode designed to safely handle higher currents. This condition is formally represented by the property \((V=4.2) \release_{[2,10]} (I<10)\), which indicates that \textit{the current will not exceed $10$ within a timeframe of $2$ to $10$ seconds, unless the voltage reaches $4.2$ volts earlier}.

    \begin{figure}[t]
        \centering
        \includegraphics[width=.76\linewidth, trim={0cm 0cm 18cm 0cm}, clip]{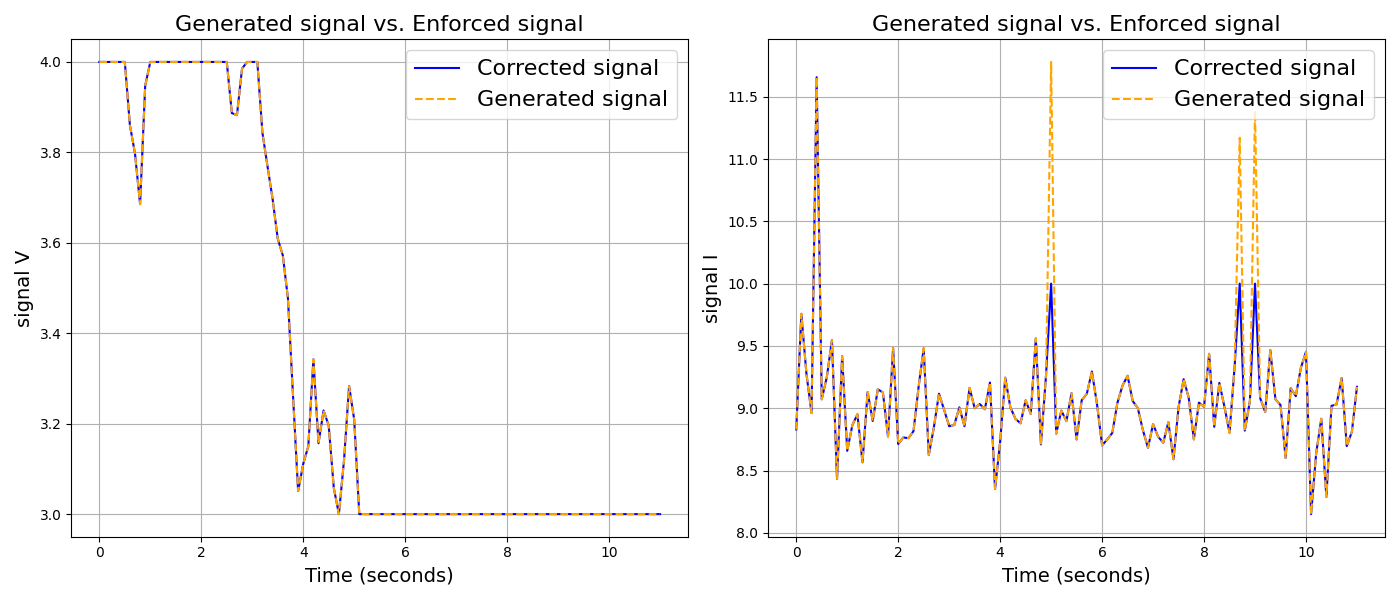}
        \includegraphics[width=.76\linewidth, trim={17.6cm 0cm 0cm 0cm}, clip]{figures/battery_corrected.png}
        \caption{Enforcement of Voltage and Current Signals against Safe Charge Property}
        \label{fig:battery}
    \end{figure}

    The results of our experiment are depicted in \cref{fig:battery}, where the blue signal represents the output after enforcement, and the orange signal is the original one. These results illustrate that during the interval from $2$ to $10$ seconds, the current $I$ is minimally adjusted to remain below $10$, provided that the voltage $V$ does not reach $4.2$ volts.

    \paragraph{Safe deceleration of AVs}
    Consider a scenario of coordinated deceleration for stability and safety in AVs, where both wheels and motor controls slow down together, helping avoid sudden stops or imbalances. It is a dual-redundant safety feature: The wheel subsystem must ensure that its value does not exceed 30 within the timeframe and ultimately reaches zero between 5 and 10 seconds. Simultaneously, the motor control subsystem has the same requirement. This condition is formally represented by the property $(w \leq 30)\until_{[5,10]} (w = 0) \land (m \leq 30)\until_{[5,10]} (m = 0)$, which indicates that \textit{both the wheels and motor control must ultimately reach $0$ within a time frame of $5$ to $10$ seconds while maintaining the values no larger than $30$ until then}.
    \begin{figure}[t]
        \centering
        \includegraphics[width=0.83\linewidth, trim={0cm 0.5cm 0cm 0.9cm}, clip]{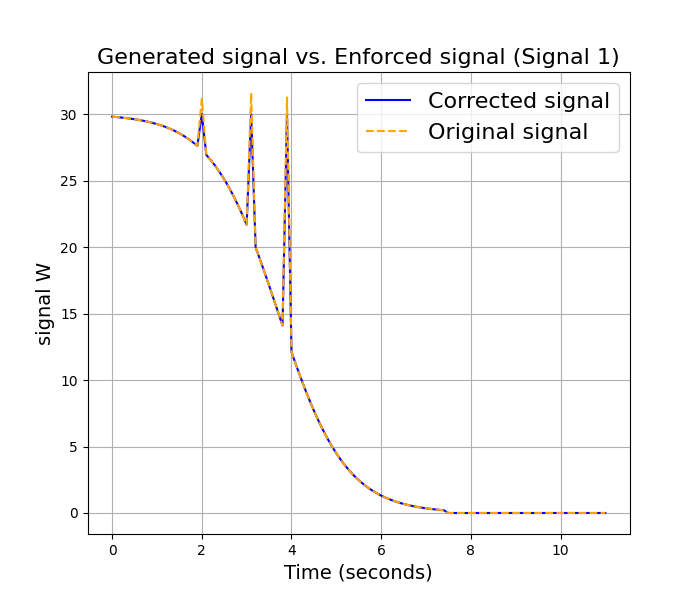}
        \includegraphics[width=0.83\linewidth, trim={0cm 0.5cm 0cm 0.6cm}, clip]{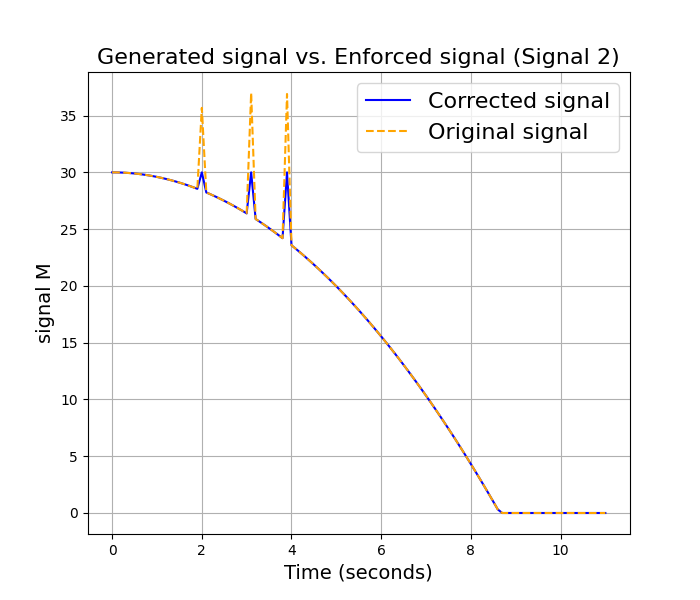}
        \caption{Enforcement of Wheel and Motor Control Signals against Safe Deceleration Property}
        \label{fig:wheel_motor}
    \end{figure}

    The results of our experiment are depicted in \cref{fig:wheel_motor}. These results illustrate that during the interval from $5$ to $10$ seconds, the wheel ($w$) and motor control ($m$) are minimally adjusted to remain below $30$, provided that these signals do not reach $0$ volts.

\subsection{Scalability Evaluation}\label{sec:scala}
    To assess the scalability of our approach, we conducted experiments in which we progressively increased the complexity of the signal - specifically, the number of violation points in the signal - to examine how enforcement time is affected. The results, presented in \cref{tab:performance} for the three scenarios mentioned earlier, show that the enforcement time (measured in milliseconds) increases in a piecewise linear fashion as the number of violations grows. This behavior is consistent with the predictions of our complexity analysis.
    \begin{table*}[t]
        \centering
            \captionsetup{font={small}}
            \caption{Experimental results with varying violation points in the signal}
            \vspace{-0.32cm}
            \label{tab:performance}
            \begin{center}
                \begin{tabular}{@{}ccccccccc@{}}
                    \toprule
                    \multirow{2}{*}{$\#v$} & \multicolumn{2}{c}{\textit{Safe stopping of AVs}} &~& \multicolumn{2}{c}{\textit{Safe charging of AVs}} &~& \multicolumn{2}{c}{\textit{Safe deceleration of AVs}} \\
                    \cmidrule{2-3} \cmidrule{5-6} \cmidrule{8-9}
                                       & $\textsf{len}(\tword)$        & \textsf{time}(ms)       &~& $\textsf{len}(\tword)$        & \textsf{time}(ms)         &~& $\textsf{len}(\tword)$        & \textsf{time}(ms)          \\ \midrule
                    2                  & 6        & 0.117     &~& 5        & 0.086      &~& 9        & 0.221      \\ 
                    4                  & 8        & 0.124     &~& 7        & 0.146      &~& 13       & 0.369      \\ 
                    6                  & 10       & 0.14      &~& 9        & 0.156      &~& 17       & 0.419      \\ 
                    8                  & 12       & 0.156     &~& 11       & 0.173      &~& 20       & 0.629      \\ 
                    10                 & 14       & 0.186     &~& 13       & 0.19       &~& 21       & 0.631      \\ 
                    12                 & 14       & 0.202     &~& 15       & 0.199      &~& 26       & 0.719      \\ 
                    14                 & 18       & 0.237     &~& 14       & 0.215      &~& 27       & 0.8        \\ 
                    16                 & 16       & 0.217     &~& 17       & 0.262      &~& 28       & 0.867      \\ 
                    18                 & 19       & 0.237     &~& 19       & 0.265      &~& 35       & 0.916      \\ 
                    20                 & 22       & 0.287     &~& 17       & 0.242      &~& 30       & 0.988      \\ \bottomrule
                    \end{tabular}
            \end{center}
            \small{ 
                $\textsf{len}(\tword)$: the length of time word encoded from the signal;~
                $\#v$: the number of violation points in signal
            } 
        \end{table*}

        Overall, our method demonstrates robust capabilities in runtime enforcement for signals against properties specified using STL. It ensures compliance with requirements for soundness, transparency, and minimal modification across all scenarios. Moreover, it exhibits high effectiveness in managing complex signals, indicating that the time required is minimal. 

\section{Related Works}
\label{sec:RL}
\paragraph{Runtime enforcement for reactive systems.}
A framework to synthesize enforcers for reactive systems, called shields, from a set of safety properties was introduced in \cite{10.1007/978-3-662-46681-0_51}. The uni-directional shield observes inputs from the environment and outputs from the system (program) and transforms erroneous outputs. It considered untimed properties expressed as automata. 

Authors in \cite{10.1145/3126500,10.1145/3092282.3092291,10.1109/TII.2019.2945520} extendes \cite{10.1007/978-3-662-46681-0_51} and considered bi-directional runtime enforcement for reactive systems. The enforcer presented a monitoring framework which monitors both the inputs and the outputs of a synchronous program and (minimally) edits erroneous inputs/outputs in order to guarantee a given property.  In \cite{10.1145/3092282.3092291,10.1109/TII.2019.2945520}, the properties are discrete properties, expressed using a variant of timed automata  
called Discrete Timed Automata (DTA) and Valued Discrete Timed Automata (VDTA). These are TAs with integer-valued clocks (i.e., FSMs extended with a set of integer variables that are used as discrete clocks, for instance, to count the number of ticks before a certain event occurs).  The use of DTA/VDTA over TA is primarily motivated by the fact that the approach can directly use a formulation similar to synchronous languages, where time is discretized. This makes the overall algorithm simple and does not require region or zone graph construction. All transitions take one tick relative to the ticks of a synchronous global clock inspired by synchronous languages. The environmental inputs are captured 
and are made available. 
During a tick, all three components – the environment, the program, and the enforcer- are executed once.

The monitoring frameworks in \cite{10.1145/3092282.3092291,10.1109/TII.2019.2945520} are for discrete systems where they sampled the execution (i.e. the inputs signal occurring in the environment) to contain a number of observable state changes. For continuous timed systems, however, variables can change arbitrarily fast. For monitoring signals of time-continuous systems for dense-time properties using \cite{10.1145/3092282.3092291,10.1109/TII.2019.2945520}, the observations can be made only at discrete moments, each observation contains only partial information. This is not effective. Because only the whole set of possible observations of a particular execution can restore all information on that execution, thus, this work contributes an enforcement mechanism for dense-time real-valued signals for continuous timed systems. 

\paragraph{STL for specifying properties of CPS} STL \cite{10.1007/978-3-540-30206-3_12} is used for specifying linear-time properties of continuous real-valued signals. 
The logic of STL is based on a bounded subset of the real-time logic MITL \cite{10.1145/227595.227602} i.e. MITL$_{[a,b]}$. 
In MITL$_{[a,b]}$ all temporal modalities are restricted to intervals of the form $[a, b]$
with $0 \leq a < b$ and $a, b \in \mathbb{Q}_{\geq 0}$, where the behavior of a system is observed for a finite time interval. 

There exist frameworks for monitoring STL properties. For example, the framework in \cite{10.1007/978-3-540-30206-3_12} automatically creates property monitors that can check whether a given signal of bounded length and finite variability satisfies the property. However, this was for offline monitoring and not for correcting the signal if not according to the property. Authors in \cite{sun2024redriver} attempt enforcement (correcting a signal) specifically for the self-driving realm. 
It is based on the predictive environment constructed by the sensors of the car. If the AV is predicted to potentially violate them in the near future (based on the quantitative semantics of STL), the REDriver framework repairs the trajectories using a gradient-driven algorithm. However, in some situations, the enforcer may not have access to the prediction of future signals.

Our work presents a more general approach to enforcing STL properties. Unlike existing literature, it adopts a more formal enforcement method, where the enforcer corrects the signal while adhering to some critical constraints.

\section{Conclusions and Future Works}
\label{sec:Conclusions and Future Works}
In this work, we developed a framework for the runtime enforcement against STL formula. This framework inputs a signal and outputs a minimally modified signal that satisfy the formula. Specially, given an STL formula, we derive timed transducers for the atomic components, compose them according to the formula, and apply them to the input timed words, which are obtained by encoding the signal. We present detail procedure for signal encoding, translating STL temporal operators into timed transducers, and an enforcement algorithm. Our approach effectively enforces a signal against an STL property on CPS.

As in \cite{10.1145/3126500,10.1145/3092282.3092291,10.1109/TII.2019.2945520}, we plan to extend the work to accommodate bidirectionality and also extend the framework for more general STL formulas.




\bibliographystyle{ACM-Reference-Format}
\bibliography{reference}

\appendix

\section*{Appendix}

\section{Proof of Propositions}\label{sec:appendix}

\restateUntil*

    \begin{proof}To prove the above proposition, it suffices to demonstrate it in two steps/implications. Once both implications are established, it follows that the proposition holds. Thus, let us prove the proposition in 2 steps:

    \begin{tcolorbox}[boxrule=.5pt,colback=white,colframe=black!75]
    \[
        1.~~~~ \llangle\automaton_{\until}\rrangle(\tword) = \bm{\omega}_{\top} \implies \signal \models p_1 \until_{[t_1, t_2]} p_2
    \]
    \end{tcolorbox}
    
    Let us consider the following cases based on the events received in time intervals: $[0, t_1], (t_1,t_2]$.
    \begin{enumerate}
        \item Case 1: when the time interval is $[0,t_1]$:\\
        Based on the events received at $t\in [0,t_1]$, we have the following sub-cases:
        \begin{enumerate}
            \item Case 1a: $p_1$ is continuously received by the transducer at $t \in [0,t_1]$ with the output of all transitions being $\top$ and $p_2$ is received at $t\in[t_1,t_1]$ with the output of the transition being $\top$ again.

            The sequence of locations visited by the transducer for this case will be $l_0, l_1, l_2$. We see that the final location $l_2$ is reached by the transducer, thus $\llangle\automaton_{\until}\rrangle(\tword) = \bm{\omega}_{\top}$.

            According to the semantics of STL, if $p_2$ is received at $t\in[t_1, t_1]$ until that if $p_1$ is continuously received, then the STL formula is satisfied by the signal corresponding to $\tword$. Thus the proposition holds for this sub-case.
            
            \item Case 1b: $p_1$ is not true at $t \in [0,t_1]$ with the output of all transitions being $\bot_{p_1}$ and $p_2$ is true at $t\in[t_1,t_1]$.

            The proposition trivially holds here as well.\\
            
        \end{enumerate}
        \item Case 2: when the time interval is $(t_1,t_2]$:\\
         Based on the events received at $t\in (t_1,t_2]$, we have the following sub-cases:
        \begin{enumerate}
            \item Case 2a: $p_2$ is received by the transducer at $t \in (t_1,t_2]$ with the output of transition being $\top$  and $p_1$ is continuously received until that with the output of all transitions being $\top$.
            
            The sequence of locations visited by the transducer for this case will be $l_0, l_1, l_3, l_2$. We see that the final location $l_2$ is reached by the transducer, thus $\llangle\automaton_{\until}\rrangle(\tword) = \bm{\omega}_{\top}$.

            According to the semantics of STL, if $p_2$ is received at $t\in (t_1,t_2]$ until that $p_1$ is continuously received, then the STL formula is satisfied by the signal corresponding to $\tword$. Thus the proposition holds for this sub-case.

            \item Case 2b: $p_2$ is not true at $t\in(t_1,t_2]$ until that time $p_1$ is true, with the output of transition being $\bot_{p_2}$. The proposition trivially holds here as well.

            \item Case 2c: $p_2$ is true at $t\in(t_1,t_2]$ however until that time $p_1$ is not true, with the output of transition being $\bot_{p_1}$. The proposition trivially holds here as well.

            \item Case 2d: $p_2$ is not true at $t\in(t_1,t_2]$ however until that time $p_1$ is also not true, with the output of transition being $\bot_{p_1}$ or $\bot_{p_2}$ accordingly. The proposition trivially holds here as well. The proposition trivially holds here as well.
             
        \end{enumerate}
    \end{enumerate}

\begin{tcolorbox}[boxrule=.5pt,colback=white,colframe=black!75]
    \[
        2. ~~~~\signal \models p_1 \until_{[t_1, t_2]} p_2  \implies \llangle\automaton_{\until}\rrangle(\tword) = \bm{\omega}_{\top}
    \]
    \end{tcolorbox}

Let us consider the following cases based on the events received in time intervals: $[0, t_1], (t_1,t_2]$.
    \begin{enumerate}
        \item Case 1: when the time interval is $[0,t_1]$: an STL formula is satisfied at $t\in [0, t_1]$ by signal $\signal$, if $p_1$ of the encoded word $\tword$ of $\signal$ is continuously true and $p_2$ of the encoded word $\tword$ of $\signal$ is true at $t\in[t_1, t_1]$.

        For that encoded word $\tword$ of signal $\signal$, the transducer makes a sequence of transitions involving locations $l_0, l_1, l_2$ (with the output being $\top$ for all the transitions) and goes to the accepting state $l_2$. Thus,  $\llangle\automaton_{\until}\rrangle(\tword) = \bm{\omega}_{\top}$ and the proposition holds.

        \item  Case 2: when the time interval is $(t_1,t_2]$: an STL formula is satisfied at $t\in (t_1,t_2]$ by signal $\signal$, if $p_2$ of the encoded word $\tword$ of $\signal$ is received at $t\in (t_1,t_2]$ and  $p_1$ of the encoded word  $\tword$ of $\signal$  is continuously true until that.

        For that encoded word $\tword$ of signal $\signal$, the transducer makes a sequence of transitions involving locations $l_0, l_1, l_3, l_2$ (with the output being $\top$ for all the transitions) and goes to the accepting state $l_2$. Thus,  $\llangle\automaton_{\until}\rrangle(\tword) = \bm{\omega}_{\top}$ and the proposition holds.\\
    \end{enumerate}
\end{proof}


\restateComposition*
\begin{proof}
    Let us prove this proposition using induction on the predicates. There will be two distinct cases based on $op\in \{\land, \lor\}$. Let us prove this proposition for $op= \{\land\}$. Similar proof will follow for $op= \{\lor\}$.\\
    
    \noindent \textit{Induction basis.} Consider STL formula $\varphi_1 \land \varphi_2$. Let us consider following cases:

    \begin{enumerate}
        \item Case 1: $\phi_1 \until_{[t_1, t_2]} \phi_2 \land \top$ \\
        where $\varphi_1 \equiv \phi_1 \until_{[t_1, t_2]} \phi_2$ and $\varphi_2 \equiv \top$.\\
        
        (Similar proof will follow for  $\top \land \phi_1 \until_{[t_1, t_2]} \phi_2$.)\\
        
        $\top$ (the true predicate) represents a property or a transducer that is always true, regardless of time constraints or inputs. A timed transducer for $\top$ would have only one state, an "accepting" state with a self-loop transition on this state allowing any input or no input to be processed at any time and $\top$ as output. 

        The structure (states and transitions) of $\land$-product of transducers $\automaton_{\phi_1 \until_{[t_1, t_2]} \phi_2}$ and $\automaton_{\top}$ will be the similar as transducer $\automaton_{\phi_1 \until_{[t_1, t_2]} \phi_2}$ (with $\top$ also an output for all transitions indicating predicate $\top$ is true).

        And from proposition \ref{propo1}, we will have following results:        
        $\llangle\automaton_{\phi_1 \until_{[t_1, t_2]} \phi_2} ~\times_{\land} ~\automaton_{\top}\rrangle(\tword) = \bm{\omega}_{\top} \iff \signal \models \phi_1 \until_{[t_1, t_2]} \phi_2 ~\land ~\top$.
        Thus, the proposition holds.
        
        \item Case 2: $p_1 \release_{[t_1, t_2]} p_2 \land \top$ \\
        (or similarly, $\top \land p_1 \release_{[t_1, t_2]} p_2$)\\

        Similar proof follows for this case as well.
    \end{enumerate}

    \noindent \textit{Induction Step.}
    \begin{enumerate}
        \item Case 1: $\phi_1 \until_{[t_1, t_2]} \phi_2 \land \phi_3 \until_{[t_3, t_4]} \phi_4$\\
        where $\varphi_1 \equiv \phi_1 \until_{[t_1, t_2]} \phi_2$ and $\varphi_2 \equiv \phi_3 \until_{[t_3, t_4]} \phi_4$.

        Let us consider sub-cases based on time intervals.
        \begin{enumerate}
            \item Case: when the time interval is $[0,t_1]$:
            \begin{enumerate}
                \item Case: $\phi_1$ and $\phi_3$ are continuously received by the transducer at $t \in [0,t_1]$ and  $t \in [0,t_3]$ respectively, with the output of all transitions being $\top$. $\phi_2$ and $\phi_4$ are received at $t\in[t_1,t_1]$ and  $t\in[t_3,t_3]$ respectively with the output of the transition being $\top$ again.
    
                The sequence of locations visited by the transducer $\automaton_{\phi_1 \until_{[t_1, t_2]} \phi_2} \times_{\land} \automaton_{\phi_3 \until_{[t_3, t_4]} \phi_4}$ for this case will be $(l_0, l_0'), (l_1, l_1'), (l_2,l_2')$ where $\{l_0, l_1, l_2\} \in L $ of $ \automaton_{\phi_1 \until_{[t_1, t_2]} \phi_2}$ and $\{l_0', l_1', l_2'\} \in L'$ of $\automaton_{\phi_3 \until_{[t_3, t_4]} \phi_4}$. We see that the final location $(l_2,l_2')$ is reached, thus $\llangle\automaton_{\automaton_{\phi_1 \until_{[t_1, t_2]} \phi_2} \times_{\land} \automaton_{\phi_3 \until_{[t_3, t_4]} \phi_4}}\rrangle(\tword) = \bm{\omega}_{\top}$.
    
                This is inline with the semantics of STL. Thus,
                $\llangle\automaton_{\varphi_1}\times_{op}\automaton_{\varphi_2}\rrangle(\tword) = \bm{\omega}_{\top} \implies \signal \models \varphi_1\,op\,\varphi_2$. 
                
                Similarly, following step 2 of proof of  proposition \ref{propo1} proof of $\signal \models \varphi_1\,op\,\varphi_2  \implies \llangle\automaton_{\varphi_1}\times_{op}\automaton_{\varphi_2}\rrangle(\tword) = \bm{\omega}_{\top} $ will follow. Thus, the proposition holds.
                
                \item Case: $\phi_1$ and $\phi_3$ is not true at $t \in [0,t_1]$ and $t \in [0,t_3]$ respectively with the output of all transitions being $\bot_{\phi_1}$ and $\bot_{\phi_3}$.  $\phi_2$ and $\phi_4$ is true at $t\in[t_1,t_1]$ and $t\in[t_3,t_3]$ respectively.
    
                The proposition trivially holds here.\\
        \end{enumerate}

            \item Case: when the time interval is $(t_1,t_2]$:\\
            Based on the events received at $t\in (t_1,t_2]$, we have the following sub-cases:
            \begin{enumerate}
                \item Case: $\phi_2$ and $\phi_4$ are received at $t \in (t_1,t_2]$ and $t \in (t_3,t_4]$ respectively with the output of transition being $\top$  and $\phi_1$ and $\phi_3$ are continuously received until that, with the output of all transitions being $\top$.
                
                The sequence of locations visited by the transducer for this case will be $(l_0, l_0'), (l_1,l_1'), (l_3,l_3'), (l_2,l_2')$. Thus, we see that the final location $(l_2,l_2')$ is reached, thus $\llangle\automaton_{\automaton_{\phi_1 \until_{[t_1, t_2]} \phi_2} \times_{\land} \automaton_{\phi_3 \until_{[t_3, t_4]} \phi_4}}\rrangle(\tword) = \bm{\omega}_{\top}$.
    
                This is inline with the semantics of STL. Thus the proposition holds.
    
                \item Case: $\phi_1$ and $\phi_3$ is true at $t \in (t_1,t_2]$ and $t \in (t_3,t_4]$ respectively. $\phi_2$ and $\phi_4$ are not true at $t\in(t_1,t_2]$ and $t \in (t_3,t_4]$ respectively with the output of transition being $\bot_{p_2}$ and $\bot_{p_4}$.
    
                The proposition trivially holds here.\\
    
                \item Case: $\phi_1$ and $\phi_3$ are not true at $t\in(t_1,t_2]$. $\phi_2$ and $\phi_4$ are also not true until that. The output of all transitions being $\bot_{p_1}$, $\bot_{p_2}$, $\bot_{p_3}$ or $\bot_{p_4}$.
    
                 The proposition trivially holds here.\\

                 \item Case: $\phi_1$ and $\phi_3$ are not true at $t\in(t_1,t_2]$. However, $\phi_2$ and $\phi_4$ are true until that. The output of all transitions being $\bot_{p_1}$ or $\bot_{p_2}$.
                \end{enumerate} 
        \end{enumerate}
        
        \item Case 2: $p_1 \release_{[t_1, t_2]} p_2 \land p_3 \release_{[t_3, t_4]} p_4$\\
        where $\varphi_1 \equiv p_1 \release_{[t_1, t_2]} p_2$ and $\varphi_2=p_3 \release_{[t_3, t_4]} p_4$

        \item Case 3: $p_1 \release_{[t_1, t_2]} p_2 \land p_3 \until_{[t_3, t_4]} p_4$\\
        where $\varphi_1 \equiv p_1 \release_{[t_1, t_2]} p_2$ and $\varphi_2\equiv  p_3 \until_{[t_3, t_4]} p_4$

        \item Case 4: $p_1 \until_{[t_1, t_2]} p_2 \land p_3 \release_{[t_3, t_4]} p_4$\\
        where  $\varphi_1 \equiv p_1 \until_{[t_1, t_2]} p_2$ and $\varphi_2\equiv \varphi_1  p_3 \release_{[t_3, t_4]} p_4$

        Cases 2, 3 and 4 can be proved similarly.
    \end{enumerate}

\end{proof}

\end{document}